\documentclass[a4paper,fleqn,usenatbib]{mnras}
\usepackage{longtable}
\usepackage{newtxtext,newtxmath}
\usepackage[T1]{fontenc}
\usepackage{ae,aecompl}
\usepackage{graphicx}	
\usepackage{amsmath}	
\usepackage{amssymb}	
\defcitealias{sfd}{SFD}
\defcitealias{barker2018}{BP18}
\defcitealias{ccm89}{CCM89}
\defcitealias{an2009}{APM}

\title[Isochrone fitting of NGC\,5904]
{Isochrone fitting of Galactic globular clusters -- I. NGC\,5904}

\author[G. A. Gontcharov, A. V. Mosenkov and M. Yu. Khovritchev]{
George A. Gontcharov,$^{1,2}$\thanks{E-mail: george.gontcharov@tdt.edu.vn}
Aleksandr V. Mosenkov,$^{3}$
and Maxim Yu. Khovritchev$^{3,4}$
\\
$^{1}$Department for Management of Science and Technology Development, Ton Duc Thang University, Ho Chi Minh City, Vietnam\\
$^{2}$Faculty of Applied Sciences, Ton Duc Thang University, Ho Chi Minh City, Vietnam\\
$^{3}$Central Astronomical Observatory, Russian Academy of Sciences, 65/1 Pulkovskoye chaussee, St. Petersburg 196140, Russia\\
$^{4}$St. Petersburg State University, 7/9 Universitetskaya nab., St. Petersburg 199034, Russia
}

\date{Accepted XXX. Received YYY; in original form ZZZ}

\pubyear{2019}

\begin{document}
\label{firstpage}
\pagerange{\pageref{firstpage}--\pageref{lastpage}}
\maketitle

\begin{abstract}
We present new isochrone fits to colour-magnitude diagrams of the Galactic globular cluster NGC 5904 (M5).
We utilise 29 photometric bands from the ultraviolet to mid-infrared by use of the data from the
{\it Hubble Space Telescope}, {\it Gaia} DR2, {\it Wide-field Infrared Survey Explorer},
Sloan Digital Sky Survey (SDSS), and other photometric data.
In our isochrone fitting we use the PAdova and TRieste Stellar Evolution Code,
the MESA Isochrones and Stellar Tracks,
the Dartmouth Stellar Evolution Program, and a Bag of Stellar Tracks and Isochrones
both for the solar-scaled and enhanced He and $\alpha$ abundances
with a metallicity about [Fe/H]$=-1.33$ adopted from the literature.
All tools  provide us with estimates of the distance, age, and extinction law to the cluster.
The best-fit distance, true distance modulus,
and age are $7.4\pm0.3$ kpc, $14.34\pm0.09$ mag, and $12.15\pm1.00$ Gyr, respectively.
The
derived distance agrees with the literature, including the {\it Gaia} DR2 parallax
with its known global zero-point correction. 
All the data and models, except some UV and SDSS data, agree with the extinction law
of Cardelli-Clayton-Mathis with $R_\mathrm{V}=3.60\pm0.05$ and $A_\mathrm{V}=0.20\pm0.02$ mag.
This extinction is twice as high as generally accepted due to a rather high extinction between 625 and 2000 nm.
An offset of the model colours instead of the high extinction in this range is a less likely, yet possible
explanation of the discovered large deviations of the isochrones from the data.
\end{abstract}


\section{Introduction}
\label{intro}

Galactic globular clusters (GCs) have been considered as an ideal laboratory to verify and calibrate the stellar
evolution theory and to study the evolution of the Galaxy in general.
Despite the fact that during the last two decades some evidence for multiple populations in most GCs became available
\citep{gratton2004},
many GCs have a dominant stellar population which is convenient for isochrone modelling and deriving some
information about the cluster.

Theoretical stellar evolution models allow one to determine the age, distance, metallicity, and reddening to the cluster.
Modern observations have given rise to an abundance of new results regarding the reliable isochrone fitting
simultaneously in all possible ultraviolet (UV), optical and infrared (IR) bands for all stages of the stellar evolution,
such as
the main sequence (MS), its turn-off (TO),  the subgiant branch (SGB), red giant branch (RGB), horizontal branch (HB),
and asymptotic giant branch (AGB).
On the one hand, new accurate photometry of individual stars in many GCs has been provided recently by
the {\it Hubble Space Telescope (HST)} from the Wide Field and Planetary Camera 2 (WFPC2)
\citep{layden2005}\footnote{\url{http://cdsarc.u-strasbg.fr/viz-bin/Cat?J/ApJ/632/266}},
Wide Field Camera 3 (WFC3) UV Legacy Survey of Galactic Globular Clusters
\citep{piotto2015, soto2017}\footnote{\url{http://groups.dfa.unipd.it/ESPG/treasury.php}}
and
Advanced Camera for Surveys (ACS) survey of Galactic globular clusters
\citep{sarajedini2007}\footnote{\url{https://www.astro.ufl.edu/~ata/public_hstgc/}},
the {\it Wide-field Infrared Survey Explorer (WISE)}
\citep{wise}\footnote{\url{http://irsa.ipac.caltech.edu/Missions/wise.html}},
the {\it Gaia} DR2
\citep{gaiabrown}\footnote{\url{https://www.astro.rug.nl/~ahelmi/research/dr2-dggc/}},
the Sloan Digital Sky Survey (SDSS) \citep{sdss},
the United Kingdom Infrared Telescope Infrared Deep Sky Survey (UKIDSS)
\citep{ukidsshewett, ukidsshambly, ukidsshodgkin}\footnote{\url{http://www.ukidss.org}},
the Panoramic Survey Telescope and Rapid Response System (Pan-STARRS) \citep{bernard2014},
and other projects.
On the other hand, modern theoretical models of the stellar evolution have been adopting to non-scaled-solar abundances,
with enhanced helium and $\alpha$-elements which are typical for GCs.

As a result, new, more accurate isochrone fitting in many photometric bands has revealed new issues.
For example, \citet[][hereafter BP18]{barker2018} have shown that `three stellar evolution models available with
{\it HST}/WFC3 and {\it HST}/ACS bandpasses all produce good fits in visible and IR bands.
However, all of them fail to match GC color-magnitude diagram (CMD) morphology in the UV.'
\citetalias{barker2018} used extinction values for each of the filters, i.e. the wavelength-extintion dependance
(or the extinction law, or the reddening curve) from \citet{schlaflyfinkbeiner2011}.
In this case, a precise theoretical model with a precise extinction law would provide the same dereddened
distance modulus, metallicity, He abundance, $\alpha$-enhancement, and age, but different extinction
for each band.
However, the results of \citetalias{barker2018} do not confirm this.
\citetalias{barker2018} attributed the discrepancies of their results to some imperfection of the models used.
Yet, this may manifest that some real extinction laws to the GCs, which were considered by \citetalias{barker2018},
may differ from the law of \citet{schlaflyfinkbeiner2011}.
We note that the `standard' extinction law of \citet[][hereafter CCM89]{ccm89} with the extinction-to-reddening ratio
$A_\mathrm{V}/E(B-V)\equiv R_\mathrm{V}=3.1$, which is typically engaged in the calculation of isochrones,
may also be far from reality in some cases.

Thus, isochrone fitting of a multi-band photometry based on some accurate estimates of metallicity, He abundance and
$\alpha$-enhancement from spectroscopy and providing some convergent values of distance and age, as well as some
extinction values for a set of wavelengths,
seems to be a good method to verify the theoretical stellar evolution models and to find the best extinction laws in
the lines of sights to GCs.
For example, \citet{hendricks2012} have used a combination of broadband near-infrared and optical
Johnson-Cousins photometry to study the extinction law along the line of sight to the GC M4,
assuming the validity of the \citetalias{ccm89} extinction law with $R_\mathrm{V}$ as a derived parameter.
The five-band photometry, which they used in their study, provides enough information to derive $R_\mathrm{V}$.
However, these data seem to be not enough to reveal a difference between a real extinction law to M4 and
the one from \citetalias{ccm89}.
In this paper our ambition is to derive an extinction law to a GC as real as possible,
using a much more extended set of photometric bands than before.

The aim of our study is to fit CMDs for the GC NGC\,5904 (M5), for which a photometry in different bands
(UV, optical, near-IR, and mid-IR) is available. With a fixed metallicity, He abundance, $\alpha$-enhancement,
we are about
to derive the most probable estimates of the age and distance to this GC and draw a realistic extinction law to it.
Also, we will estimate the accuracy of the isochrones under consideration.

NGC\,5904 is a GC at RA(2000)$=15^h18^m33^s$ and DEC(2000)$=+2\degr04\arcmin52\arcsec$ or 
$l=3.859\degr$ and $b=+46.796\degr$.
The most comprehensive database of GCs by \citet{harris}\footnote{\url{https://www.physics.mcmaster.ca/~harris/mwgc.dat}},
revision of 2010,
provides a distance to it of 7.5\,kpc, reddening $E(B-V)=0.03$\,mag, [Fe/H]$=-1.29$, and
apparent visual distance modulus $(m-M)_\mathrm{V}=14.46$.
Besides the wealth of available photometry, NGC\,5904 is selected by us as a GC with no strong evidence for multiple
CMD sequences,
with accurate spectroscopic estimates of its metallicity, He abundance, and $\alpha$-enhancement,
with low foreground and differential reddening, but with quite discrepant estimates of the foreground reddening
from various data sources.

Rather different estimates of $E(B-V)$ for NGC\,5904 include:
\begin{itemize}
\item $0.00\pm0.02$ \citep{green2015},
\item $0.02\pm0.04$ \citep{arenou},
\item $0.03\pm0.02$ \citep{2015ApJ...798...88M},
\item $0.03\pm0.02$ \citep{schlaflyfinkbeiner2011},
\item $0.04\pm0.03$ \citep{drimmel}\footnote{The estimates, based on the \citet{drimmel} map, are calculated by 
use of the code of \citet{bovy2016} \url{https://github.com/jobovy/mwdust}},
\item $0.04\pm0.03$ \citep{lallement2018}\footnote{\url{http://stilism.obspm.fr}},
\item $0.04\pm0.03$ \citep[][hereafter SFD]{sfd},
\item $0.07\pm0.02$ \citep{green2018}\footnote{\url{http://argonaut.skymaps.info/}},
\item $0.10\pm0.04$ \citep{gould, av}, and
\item $0.12\pm0.04$ \citep{g17}.
\end{itemize}

These estimates are not independent. For example, \citet{2015ApJ...798...88M}, \citet{schlaflyfinkbeiner2011},
\citet{drimmel}, and \citet{lallement2018} have similar reddening zero points by use of the same data from
\citetalias{sfd} or by use of \citetalias{sfd} as the prior for a poor dataset at such a high latitude ($b\approx47\degr$).
More details on this topic are presented by \citet{gm2017big, gm2018, gm2019}.

These reddening estimates originate from observational results for very different wavelengths, which were
inter- or extrapolated to $E(B-V)$ mostly by use of the \citetalias{ccm89} extinction law with $R_\mathrm{V}=3.1$.
In the case of other extinction law for the dust medium between us and NGC\,5904,
this inter- or extrapolation may give such discrepant estimates of $E(B-V)$.
For example, we calculate here the values $E(B-V)$ for \citet{av} and \citet{g17} from their original estimates
$E(J-K_s)=0.056\pm0.02$ and $0.066\pm0.02$~mag, respectively, assuming the extinction coefficient
$E(J-K_s)/E(B-V)=0.533$ for our stars (mostly with $5000<T_\mathrm{eff}<7000$ K) from \citet{casagrande2014}.
However, an increase of this extinction coefficient with another extinction law
would decrease the $E(B-V)$ estimates of \citet{av} and \citet{g17} and make them closer to the others.

Yet, a deviation of the extinction law is not the only possible reason of this diversity of the reddening estimates.
In addition, we refer the reader to \citet{gm2018} who discuss some other reasons:
some systematic errors of the estimates from
\citetalias{sfd}, \citet{2015ApJ...798...88M} and their followers,
as well as a low accuracy of the estimates from \citet{arenou} due to a lack of high latitude stars in their sample.

Thus, these reddening data sources may be more or less reliable for describing the dust medium
between us and NGC\,5904. These data sources, except \citet{schlaflyfinkbeiner2011} and \citet{green2018},
have been verified by \citet{gm2017, gm2017big, gm2018} in their ability to put stars with the best {\it Gaia} DR1
Tycho--Gaia Astrometric Solution parallaxes \citep{tgas}
among the isochrones in the Hertzsprung--Russell diagram.
It has been found that the mean $E(B-V)$ at high Galactic latitudes behind the Galactic dust layer seems to be
underestimated by \citet{green2015}, \citet{2015ApJ...798...88M}, \citet{drimmel} and \citetalias{sfd}, while
best estimated by \citet{av} and \citet{g17}.
For NGC\,5904, we
see the same lineup of the estimates.
Therefore,
the case of NGC\,5904 seems to be suitable for further testing, whether lower or higher reddening estimates
at such a high latitude are more reliable.
Anyway, such a large diversity of the reddening estimates should be explained.
It is important not only for exploring the nature of GCs, but also for improving the reddening/extinction estimates.

An attention to the large uncertainty (30 per cent at best) of these estimates should be paid.
Up to now, this issue has not been considered thoroughly.
For example, \citet[][hereafter APM]{an2009} in their tables~6, 8 and 10 provided systematic errors of an isochrone
fitting for NGC\,5904. They suggested a 20 per cent error in $E(B-V)$ from \citetalias{sfd}.
Consequently, they stated that the distance uncertainty dominates in the uncertainty of the resulting
model colour, distance modulus, and age.
However, the error of the emission-to-reddening calibration of SFD is 0.028~mag \citepalias{sfd}.
Hence, this error appears to be 70 per cent instead of 20 per cent.
With this estimate it is evident from tables~6, 8 and 10 of \citetalias{an2009} that the uncertainty of the reddening
(and, consequently, of the extinction law) dominates.

The deviation of the extinction law is not the only explanation of the systematic colour offset of the isochrones 
from the observations in CMDs. It can also be explained by some errors associated with the detailed bandpass shape, 
absolute flux calibration, accepted colour-$T_\mathrm{eff}$ relation, and with other colour-dependent imperfections of 
the model isochrones,
providing some inherent offsets of the model colours 
(see \citet{clem2004, kucinskas2006, casagrande2014, casagrande2018a, choi2018, vandenberg2018}.

However, these reasons would lead to some similar offsets for different GCs.
Yet,
figures 5--8 in \citetalias{an2009} for the isochrone fitting of the CMDs show quite different offsets
for different GCs.
Hence, these discrepancies are mostly due to an inaccurate representation of some properties of these GCs.
Among such properties, the metallicity is well-estimated for these GCs, while some reasonable variations of age, He and
$\alpha$ abundance could not shift the isochrones considerably.
Therefore, only the extinction law, which they used in their study, seems to be responsible for the observed
offsets.

Nevertheless, the systematic accuracy of the models and isochrones used should be estimated.
Our approach may be fruitful in this way, since 
`the consistency of predicted colour-T$_\mathrm{eff}$ relations can be tested by determining if the same interpretation
of the data (including discrepancies between theory and observations) is found on many different CMDs' 
\citep{casagrande2018a}.
Hereafter, we refer any systematic colour offset of an isochrone from observations as the corresponding reddening,
though it can be partially or completely due to an imperfection of the isochrone.

This paper is organised as follows. We discuss the metallicity of NGC\,5904 in Sect.~\ref{metal}.
In Sect.~\ref{photo} we describe the photometry used.
In Sect.~\ref{iso} we describe different theoretical models of the stellar evolution and consequent isochrones.
We also draw some immediate conclusions from their application to the CMDs with the `canonical'
distance and reddening from \citet{harris}.
The results of our isochrone fitting are presented in Sect.~\ref{results} and discussed in Sect.~\ref{discuss}.
We summarise our main findings and conclusions in Sect.~\ref{conclusions}.

\section{Metallicity}
\label{metal}

We adopt the spectroscopic estimate of metallicity [Fe/H]$=-1.33\pm0.06$ derived for NGC\,5904 by
\citet{carretta2009c} within an abundance scale well-defined from a large, precise and homogeneous data-set of [Fe/H]
abundances of GCs. This estimate seems to obsolete the spectroscopic ones of \citet{gratton1986} ($-1.42\pm0.05$) and
\citet{carretta1997} ($-1.11\pm0.11$).
Another spectroscopic estimate [Fe/H]$=-1.36\pm0.07$ \citep{kraft2003}, based on a different abundance scale, seems
to obsolete the estimate $-1.17\pm0.01$ by \citet{sneden1992}.
Other spectroscopic estimates, such as $-1.40\pm0.06$ \citep{zinn1984}, even being on different abundance scales,
tend to be within [Fe/H]$=-1.33\pm0.10$.
This uncertainty of $\pm0.1$ dex corresponds to the uncertainty of colours of $<\pm0.01$ mag
at the TO, MS, SGB, HB, and fainter halves of the RGB and AGB, while up to $0.03$ mag at brighter halves of the RGB and AGB.
The latter uncertainty is still lower than some uncertainties of the models dominating at the brighter halves of the RGB
and AGB.
It is seen in CMDs, for example, from Fig.~\ref{f814_ks}--\ref{gw1} and \ref{best336_438}--\ref{bestzks} in the Appendix, 
where different models with the same
[Fe/H]$\approx-1.33$ provide isochrones several hundredths magnitudes as bluer, as redder the
fiducial sequence at the brighter halves of the RGB and AGB.
This is a reason to pay more attention to the TO, SGB, MS than to the RGB and AGB in our isochrone fitting.

\citet{carretta2009a} and \citet{carretta2009b} found large star-to-star abundance variations for NGC\,5904.
For example, for the majority of investigated stars they found $-0.2<$[O/Fe]$<+0.4$.
The average values are [O/Fe]$=+0.08\pm0.23$, [Na/Fe]$=+0.25\pm0.24$, [Mg/Fe]$=+0.41\pm0.07$, [Al/Fe]$=+0.27\pm0.29$, and
[Si/Fe]$=+0.30\pm0.05$.
This suggests an enhanced average $\alpha$ abundance with its large variations from star to star.
Moreover, \citet{dellagli} have provided some evidence for an enhanced He abundance of $0.25<Y<0.31$ for the stars
which they considered in NGC\,5904.
Hence, hereafter we consider both the scaled-solar and He-$\alpha$-enhanced isochrones for the same [Fe/H],
age, reddening, and distance (except for the models with the scaled-solar abundances only) to bracket the observed
scatter of the cluster abundances.
We test the difference between the scaled-solar and He-$\alpha$-enhanced isochrones and, consequently,
the importance of taking into account the He-$\alpha$-enhancement.
In the case of a large difference, we fit the isochrones to the CMD so that the bulk
of stars appear between the scaled-solar and non-scaled-solar isochrones of the same [Fe/H], age, and distance.

\begin{figure*}
\includegraphics{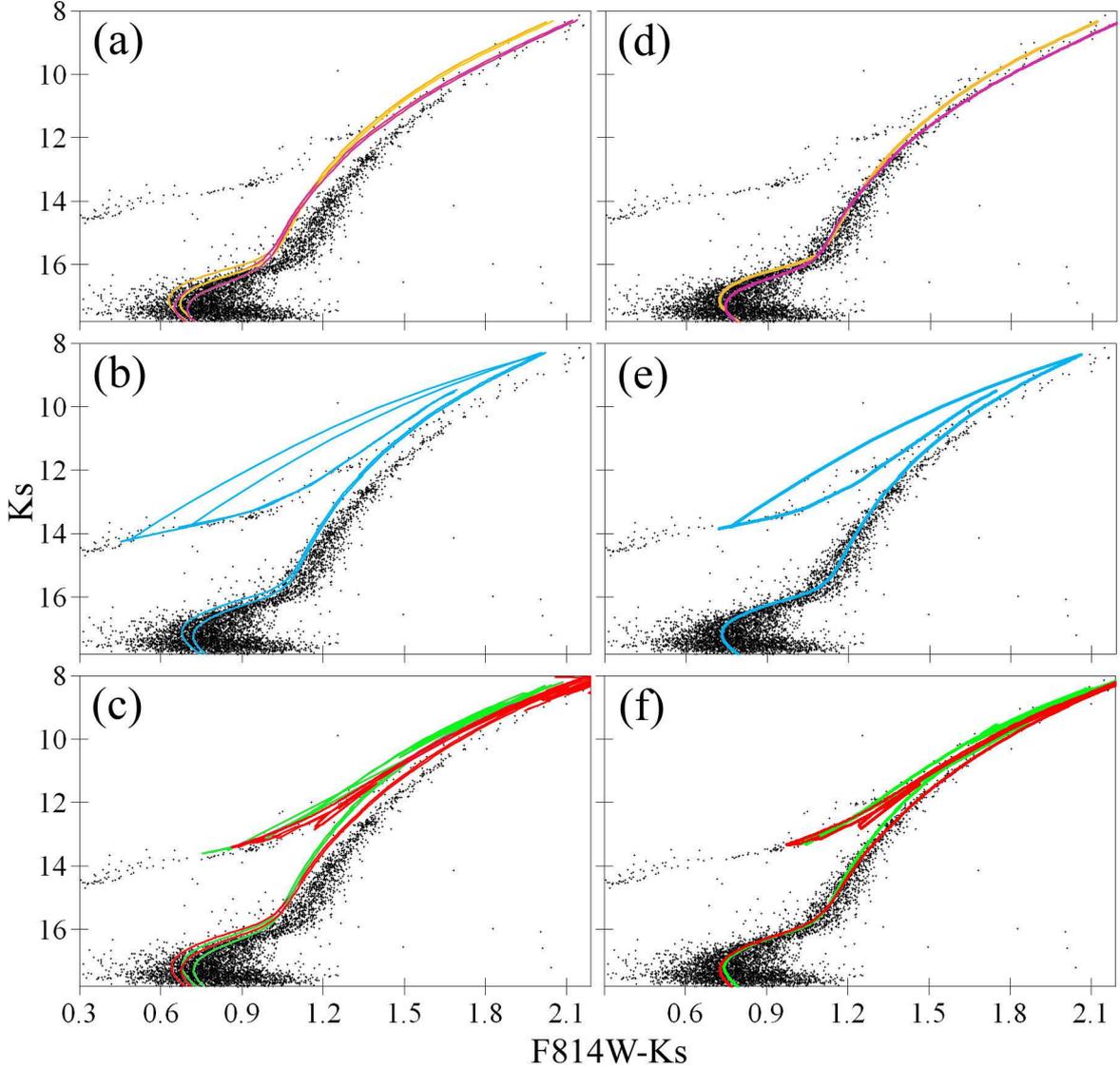}
\caption{$F814W-K_s$ versus $K_s$ CMD of NGC\,5904 with the isochrones from
(a) and (d) DSEP solar-scaled (yellow), DSEP He and $\alpha$ enhanced (magenta),
(b) and (e) new BaSTI solar-scaled (blue),
(c) and (f) PARSEC solar-scaled (green) and MIST solar-scaled (red).
The left column of the plots shows the pairs of the isochrones for 11 (left in the pair) and 13 Gyr (right in the pair)
for the distance and reddening from \citet{harris} (7.5 kpc and $E(B-V)=0.03$~mag) and the \citetalias{ccm89}
extinction law with $R_V=3.1$.
The right column of the plots shows the best fit isochrones for the age, distance and reddening from Table~\ref{fit}.
}
\label{f814_ks}
\end{figure*}

\begin{figure*}
\includegraphics{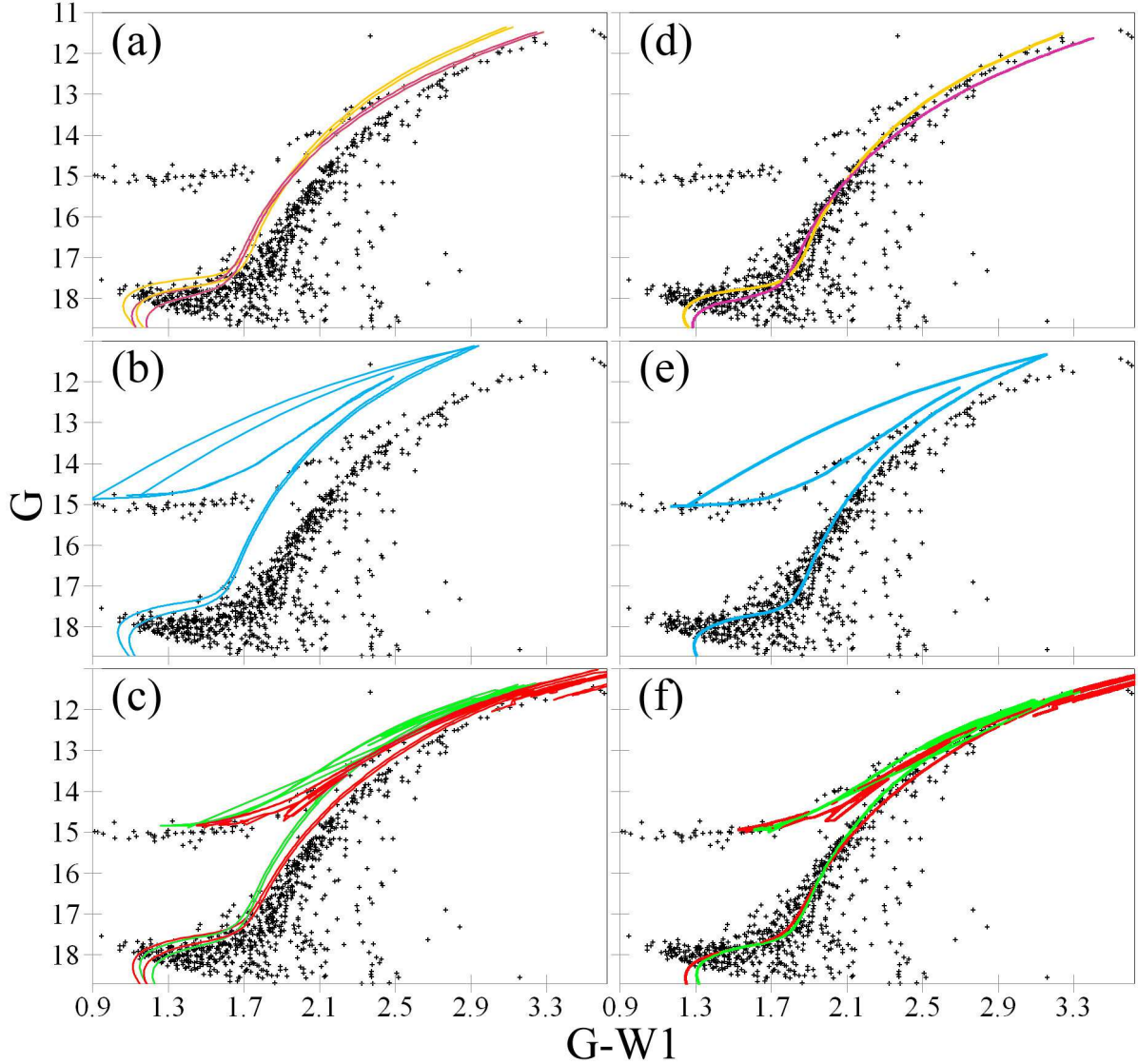}
\caption{The same as Fig.~\ref{f814_ks} but for $G-W1$ versus $G$.
}
\label{gw1}
\end{figure*}

\section{Photometry}
\label{photo}

To create CMDs for NGC\,5904, we use the following photometry in 29 filters:
\begin{itemize}
\item 11472 stars with the photometry in the $F275W$, $F336W$ and $F438W$ filters from the {\it HST} WFC3 UV Legacy Survey
of GCs \citep{piotto2015, soto2017},
\item 48860 stars with the photometry in the $F606W$ and $F814W$ filters from the {\it HST} ACS GC Survey \citep{sarajedini2007},
\item 7582 stars with the {\it HST} WFPC2 photometry in the $F555W$ and $F814W$ filters \citep{layden2005},
\item 6136 stars with the {\it Gaia} DR2 photometry in the $G$, $G_\mathrm{BP}$ and $G_\mathrm{RP}$ filters
for the stars identified as NGC\,5904 members by \citet{gaiahelmi},
\footnote{The Gaia DR2 photometry is affected by systematic errors \citep{maiz2018}.
However, in our case they are negligible.}
\item 877 stars with the {\it WISE} photometry in the $W1$ band
among the {\it Gaia} DR2 stars identified as NGC\,5904 members by \citet{gaiahelmi},
\item 3493 stars with the UKIDSS LAS photometry in the $H$ filter
among the {\it Gaia} DR2 stars identified as NGC\,5904 members by \citet{gaiahelmi},
\item from 50925 to 58591 stars (depending on the filter) with the $UBVRI$ photometry
\citep{viaux2013}\footnote{\url{http://cdsarc.u-strasbg.fr/viz-bin/Cat?J/A\%2bA/558/A12}},
\item 14379 stars with the $J$ and $K_s$ photometry obtained with SOFI at the New Technology Telescope
(NTT)\footnote{\url{http://www.eso.org/sci/facilities/lasilla/instruments/sofi/overview.html}}
and with NICS at the National Telescope Galileo
(TNG)\footnote{\url{http://www.tng.iac.es/instruments/nics/}}
\citep{coppola2011},
\item from 9088 to 27803 stars (depending on the filter) with the SDSS photometry in the $ugriz$ filters
\citep{an2008, an2009}\footnote{\url{http://classic.sdss.org/dr6/products/value_added/anjohnson08_clusterphotometry.htm},
isochrones are at \url{http://www.astronomy.ohio-state.edu/iso/sdss.html}}.
\item 17250 stars with both the Pan-STARRS and $UBVRI$ photometry,
2157 stars common to Pan-STARRS and {\it Gaia},
1510 stars common to Pan-STARRS and {\it WISE}, as well as
the fiducial sequences derived by \citet{bernard2014} in the five bands of the Pan-STARRS1 photometric system
($g_\mathrm{P1}$, $r_\mathrm{P1}$, $i_\mathrm{P1}$, $z_\mathrm{P1}$, and $y_\mathrm{P1}$).
\end{itemize}

The effective wavelengths $\lambda_\mathrm{eff}$ in nm and the median precision of the photometry
before eliminating stars with poor photometry are given for each filter in Table~\ref{filters}.
Each star has a photometry in some, not necessarily in all filters.

\begin{table}
\def\baselinestretch{1}\normalsize\normalsize
\caption[]{The effective wavelength (nm) and median precision of the photometry (mag)
for the filters under consideration.
}
\label{filters}
\[
\begin{tabular}{llcc}
\hline
\noalign{\smallskip}
 Telescope & Filter & $\lambda_\mathrm{eff}$ & Median precision \\
\hline
\noalign{\smallskip}
{\it HST}/WFC3  & $F275W$         & 274 & 0.02 \\
{\it HST}/WFC3  & $F336W$         & 329 & 0.02 \\
Various         & $U_\mathrm{Landolt}$   & 354 & 0.03 \\
SDSS            & $u$             & 360 & 0.08 \\
Various         & $B_\mathrm{Landolt}$   & 437 & 0.02 \\
{\it HST}/WFC3  & $F438W$         & 437 & 0.01 \\
SDSS            & $g$             & 471 & 0.04 \\
Pan-STARRS1     & $g_\mathrm{P1}$           & 480 & 0.01 \\
{\it Gaia} DR2  & $G_\mathrm{BP}$ & 514 & 0.04 \\
{\it HST}/WFPC2 & $F555W$         & 541 & 0.03 \\
Various         & $V_\mathrm{Landolt}$   & 547 & 0.01 \\
{\it HST}/ACS   & $F606W$         & 588 & 0.01 \\
Pan-STARRS1     & $r_\mathrm{P1}$           & 620 & 0.01 \\
SDSS            & $r$             & 621 & 0.04 \\
{\it Gaia} DR2  & $G$             & 625 & 0.01 \\
Various         & $R_\mathrm{Landolt}$   & 667 & 0.03 \\
SDSS            & $i$             & 743 & 0.05 \\
Pan-STARRS1     & $i_\mathrm{P1}$           & 746 & 0.01 \\
{\it Gaia} DR2  & $G_\mathrm{RP}$ & 762 & 0.03 \\
{\it HST}/ACS   & $F814W$         & 794 & 0.01 \\
Various         & $I_\mathrm{Landolt}$   & 812 & 0.01 \\
{\it HST}/WFPC2 & $F814W$         & 837 & 0.02 \\
Pan-STARRS1     & $z_\mathrm{P1}$           & 860 & 0.01 \\
SDSS            & $z$             & 885 & 0.10 \\
Pan-STARRS1     & $y_\mathrm{P1}$           & 960 & 0.01 \\
NTT/TNG         & $J$             & 1229 & 0.03 \\
UKIDSS          & $H$             & 1629 & 0.02 \\
NTT/TNG         & $K_s$           & 2147 & 0.04 \\
{\it WISE}      & $W1$            & 3316 & 0.05 \\

\hline
\end{tabular}
\]
\end{table}


Based on the distribution of stars by the precision of their photometry,
we select only stars with errors of less than
0.05~mag in the WFC3 and ACS colours,
0.055~mag in the WFPC2 colour,
0.05~mag in the $UBVRI$ colours,
0.06~mag in the {\it Gaia} and {\it Gaia} -- UKIDSS colours,
0.10~mag in the {\it Gaia -- WISE} colours,
0.05~mag in the $J-K_s$ colour,
0.06~mag in the SDSS colours,
and
0.05~mag in Pan-STARRS colours.

Finally, we have 5 multicolour series in the optical range:
{\it HST} (WFC3 and ACS),
{\it Gaia},
$UBVRI$,
SDSS, and
Pan-STARRS.
We have cross-identified them with each other and with the IR photometry, either with NTT/TNG $J$ and $K_s$, or UKIDSS $H$, 
or {\it WISE} $W1$.
Only the {\it HST} WFPC2 photometry from \citet{layden2005} is not cross-identified to the remaining data.
It covers only the MS and partially the SGB, thus, providing a lower accuracy of the derived age and 
distance (they are, thus, taken fixed as 12 Gyr and 7.4 kpc, respectively), yet, a fairly high accuracy of the derived 
reddening. Therefore, the WFPC2 data are used for independent control only.

These photometric data allow us to consider and fit isochrones to more than 100 CMDs with different colours
between the UV and middle IR.
Each of the five multicolour series provides us with independent results.
Additional results are derived for several IR colours.

\section{Theoretical models and isochrones}
\label{iso}

To fit the CMDs of NGC\,5904, we use the following theoretical models of the stellar evolution for creating isochrones:
\begin{itemize}
\item the PAdova and TRieste Stellar Evolution Code (PARSEC)
\citep{bressan}\footnote{\url{http://stev.oapd.inaf.it/cgi-bin/cmd}}
with [Fe/H]$=-1.35$, $Z=0.00068$, $Y=0.25$, [$\alpha$/Fe]$=0$, mass loss 0.2 and the solar $Z=0.0152$,
\item the MESA Isochrones and Stellar Tracks (MIST)
\citep{paxton2011, paxton2013, mist, choi2016}\footnote{\url{http://waps.cfa.harvard.edu/MIST/}}
with [Fe/H]$=-1.35$, $Z=0.00066$, $Y=0.25$, [$\alpha$/Fe]$=0$, rotation
$v_\mathrm{initial}/v_\mathrm{critical}=0.4$ and the solar $Z=0.0142$,
\item the Dartmouth Stellar Evolution Program (DSEP)
\citep{dotter2007, dotter2008}\footnote{\url{http://stellar.dartmouth.edu/models/}},
scaled-solar abundance with [Fe/H]$=-1.35$, $Z=0.00074$, $Y=0.246$, [$\alpha$/Fe]$=0$ and
enhanced abundance with [Fe/H]$=-1.35$, $Z=0.0013$, $Y=0.33$, [$\alpha$/Fe]$=+0.40$,
both with the solar $Z=0.0189$,
\item A Bag of Stellar Tracks and Isochrones (BaSTI) in the two versions:
the old version \citep{basti2004, basti2006, basti2013}
\footnote{\url{http://albione.oa-teramo.inaf.it}},
scaled-solar abundance with [Fe/H]$=-1.27$, $Z=0.0010$, $Y=0.246$, [$\alpha$/Fe]$=0$ and
enhanced abundance with [Fe/H]$=-1.31$, $Z=0.0019$, $Y=0.30$, [$\alpha$/Fe]$=+0.5$,
both with the solar $Z=0.019$;
the new version \citep{newbasti}\footnote{\url{http://basti-iac.oa-abruzzo.inaf.it/}}
with [Fe/H]$=-1.35$, $Z=0.000701$, $Y=0.2478$, [$\alpha$/Fe]$=0$,
overshooting, diffusion, mass loss 0.3
\footnote{See \citet{choi2018} for a discussion of an effect of variying mass loss and other parameters on CMDs.}
and the initial solar $Z=0.0172$ and $Y=0.2695$.
\item the isochrones from \citetalias{an2009}\footnote{\url{http://www.astronomy.ohio-state.edu/iso/sdss.html}}
with [Fe/H]$=-1.5$, [$\alpha$/Fe]$=+0.3$ for the SDSS filters
are used here only for testing because they draw the MS and TO only, and only for the ages 11.22 and 12.59 Gyr
(both ages give the same accuracy of the fits).
\end{itemize}

\begin{table}
\def\baselinestretch{1}\normalsize\normalsize
\caption[]{The fiducial sequence for NGC 5904 $F814W$ ({\it HST} ACS) versus $K_s$.
The complete table is available online.
}
\label{fiducial}
\[
\begin{tabular}{rr}
\hline
\noalign{\smallskip}
$F814W$ & $K_s$ \\
\hline
\noalign{\smallskip}
16.20 & 16.4 \\
16.05 & 16.2 \\
15.90 & 16.0 \\
15.75 & 15.8 \\
15.60 & 15.6 \\
\ldots & \ldots \\
\hline
\end{tabular}
\]
\end{table}


Unfortunately, the current releases of PARSEC and MIST include the scaled-solar models only.

We note that the rotating and non-rotating MIST models give indistinguishable isochrones.

For every CMD we calculate the cluster fiducial sequence defined as the locus of the number density peaks on the CMD.
This sequence represents a colour-magnitude relation for single stars of a dominant population in a cluster.
We calculate the fiducial sequence as series of the median points in 
magnitude bins that have a size which varies as a function of the number of stars and photometric errors 
from 0.1 to 0.2 mag.
This is an usual approach presented, for example, by
\citet{layden2005}, \citet{an2008}, \citet{hendricks2012}, \citet{bernard2014}, and \citet{vandenberg2018}.
An example fiducial sequence for $F814W$ ({\it HST} ACS) versus $K_s$ is presented in Table~\ref{fiducial}.
All other fiducial sequences can be provided on request.

We note that the fiducial sequences for NGC\,5904 are relatively easily defined due to
(i) a high degree of the completeness of the stellar samples under consideration, at least, between the HB and TO,
(ii) low differential reddening,
(iii) low contamination from foreground/background stars at such a high latitude, and
(iv) low percentage of stars with a different age/metallicity.

The colour of the fiducial sequence for the faintest stars depends on their distance from the cluster's center.
This may be due to a contamination of the periphery of the cluster by the field stars or due to a crowding effect
in the centre.
However, this shift of the fiducial sequence is always $<0.01$ mag, i.e. negligible.

To derive the best reddening, distance, and age, we calculate the isochrones for a grid of some reasonable 
ages (10--14 Gyr with a step of 0.5 Gyr), 
distances (6.5--8.5 kpc with a step of 0.1 kpc) 
and reddenings [between $-0.1$ mag and the value calculated from $E(B-V)=0.12$ 
(the highest estimate in Sect.~\ref{intro})
and the \citetalias{ccm89} extinction law with $R_\mathrm{V}=5$, with a step of 0.005 mag].
Then we select the isochrone with a minimal total offset between the empirical fiducial points and their theoretical 
equivalents on this isochrone. 
To pay less attention to the RGB and AGB in our isochrone fitting, as discussed in Sect.~\ref{metal},
we consider only fiducial points at the TO, SGB, HB, the brighter part of the MS 
(limited by $<21$ mag for the optical bands) and the part of the RGB fainter than the HB, 
if the isochrone and the data cover these ridges.

The distance is better constrained by the magnitudes of the HB and SGB,
the age -- by the length and slope of the SGB and by the magnitude difference between the HB and SGB,
while the reddening -- by the overall colour offset of the isochrone w.r.t. the fiducial sequence.

Some changes in each parameter resulted in noticeably better or worse fits while the other parameters remained constant.
These changes vary from colour to colour, but typically are 1~Gyr ($\approx$8 per cent) in age, 
200~pc ($\approx$3 per cent) in
distance, and 0.005~mag ($\approx$10 per cent) in reddening.
We consider these values as preliminary estimates of the uncertainties of the fitting.

\subsection{Immediate conclusions}

Some immediate conclusions for a further fitting can be made from the fitting of the isochrones with the
`canonical' distance 7.5~kpc and reddening $E(B-V)=0.03$~mag from \citet{harris} together with the \citetalias{ccm89}
extinction law with $R_\mathrm{V}=3.1$.
The CMDs $F814W-K_s$ versus $K_s$ and $G-W1$ versus $G$ appear quite informative.
They are presented in the left plots of Fig.~\ref{f814_ks} and \ref{gw1}, respectively, with the isochrones overimposed.
Each isochrone is presented as a pair for the ages 11 (left) and 13 Gyr (right).
This covers a range with the most probable estimates of the age for NGC\,5904 from the literature.
Yet, it is seen that the isochrones in each pair differ only on the TO, SGB and blue HB.
This means that, on the one hand, the age is poorly defined by the fitting, while, on the other hand,
any reasonable variations of the age could not eliminate the evident isochrone-fiducial offset.

An almost correct vertical position of the isochrones w.r.t. the HB and the SGB
shows that the accepted distance of 7.5 kpc is quite close to the truth.

Another important immediate conclusion from
Fig.~\ref{f814_ks} and \ref{gw1} is that DSEP shows
quite little (few hundredths of mag) difference between the scaled-solar and He-$\alpha$-enhanced isochrones of the
same [Fe/H], age, reddening, and distance.
The same is
evident from the next figures for all the colours, both for DSEP and old BaSTI.
We explain this by the fact that with the same [Fe/H] the $\alpha$-enhancement makes the star cooler, whereas
He-enhancement makes it hotter.
Finally, they nearly compensate each other: an example is shown by \citet[their figure 14, (b) and (d)]{viaux2013}.
Thus, we can use the scaled-solar isochrones as an acceptable description for this GC.

The most important immediate conclusion from Fig.~\ref{f814_ks} and \ref{gw1} is that the Harris's $E(B-V)=0.03$
mag gives a noticeable colour offset of all the isochrones from the bulk of the stars.
\footnote{Recently, 
\citet{casagrande2018b} have provided a correction to a systematic error in $G$ in their equation (3).
Also,
\citet{weiler2018} have presented response curves of $G$, $G_\mathrm{BP}$, and $G_\mathrm{RP}$ that 
differ from the curves previously pulished by \citet{gaiaevans}.
Up to now only PARSEC has provided the isochrones with the both kinds of the curves.
We discover that both the systematic correction of $G$ and the change of the curves 
make the $G-W1$ and $G_\mathrm{RP}-W1$ isochrones even more bluer by $0.005$ mag, hence, increasing the reddenings
$E(G-W1)$ and $E(G_\mathrm{RP}-W1)$ needed to fit the isochrones to the fiducial sequences.
Therefore, we use only the curves from \citeauthor{gaiaevans}, and $G$ without a correction.}
The same is typical for the other CMDs.
Naturally, it can be solved by accepting a higher reddening.
However, as mentioned in Sect.~\ref{intro}, this colour offset may be due to an offset
of the model colours rather than due to the underestimated reddening.
Anyway, any reasonable variation of metallicity, abundance, age, and distance cannot eliminate this offset.

Indeed, for any model and isochrone an increase of a parameter shifts the isochrone w.r.t. the fiducial sequence
as follows [see also figure 11 of \citet{choi2018}]:
\begin{itemize}
\item the increasing He abundance makes the RGB bluer, the SGB shorter, and the SGB less s-shaped;
\item the increasing $\alpha$-enhancement makes the isochrone redder and fainter;
\item the increasing age makes the SGB shorter and the HB-SGB magnitude difference larger;
\item the increasing distance makes the isochrone fainter;
\item the increasing reddening makes the isochrone redder and fainter;
\item the increasing extinction law parameter, such as $R_\mathrm{V}$, 
makes any optical-IR isochrone redder and fainter.
\end{itemize}
Hence, for a fixed reddening there are only three ways to shift an isochrone redward:
(i) a lower He abundance, or (ii) a higher $\alpha$ abundance, or (iii) a different extinction law, e.g. with 
a higher $R_\mathrm{V}$.
Yet, both the abundances are tightly constrained by the spectroscopy.
Moreover, the scaled-solar and He-$\alpha$-enhanced isochrones almost coincide.
Thus, a higher reddening, 
or another extinction law, or a model colour improvement are the only solutions.

The right plots of Fig.~\ref{f814_ks} and \ref{gw1} present the best fits with the derived
age, distance and reddening (more generally, isochrone-fiducial colour offset) given in Table~\ref{fit}.

\begin{table*}
\def\baselinestretch{1}\normalsize\normalsize
\caption{The results of the isochrone fitting for various models and colours in comparison with the same
reddenings predicted from \citet{harris}.
The values used to derive the average age and distance are highlighted with bold.
The adopted values of age and distance are highlighted with italic.
}
\label{fit}
\[
\begin{tabular}{lccccccc}
\hline
                 & PARSEC         & MIST           & DSEP           & Old BaSTI      & New BaSTI      & APM & Harris \\
\hline
$E(F275W-F336W)$ & $0.073\pm0.10$ & $0.105\pm0.08$ & $0.000\pm0.12$ & $0.066\pm0.05$ & $0.067\pm0.07$ & & 0.03 \\
age, Gyr         & 11.5           & 13.0           & 11.0           & \textbf{11.5}  & 11.5           & & \\
distance, kpc    & 7.5            & \textbf{6.9}   & 8.3            & 7.1            & \textbf{7.2}   & & \\
\hline
$E(F336W-F438W)$ & $0.060\pm0.10$ & $0.080\pm0.06$ & $0.030\pm0.03$ & $0.035\pm0.04$ & $0.040\pm0.04$ & & 0.03 \\  
age, Gyr         & 12.0           & 12.5           & \textbf{11.0}  & \textbf{12.0}  & 12.5           & & \\
distance, kpc    & 7.5            & \textbf{7.1}   & 7.5            & 7.3            & \textbf{7.5}   & & \\
\hline
$E(F438W-F606W)$ & $0.101\pm0.01$ & $0.114\pm0.04$ & $0.070\pm0.02$ & $0.089\pm0.03$ & $0.070\pm0.03$ & & 0.04 \\  
age, Gyr         & 11.5           & 13.5           & \textbf{12.0}  & \textbf{12.5}  & 13.0           & & \\
distance, kpc    & \textbf{7.3}   & \textbf{7.0}   & 7.2            & 6.8            & \textbf{7.2}   & & \\
\hline
$E(F606W-F814W)$ & $0.045\pm0.01$ & $0.050\pm0.03$ & $0.040\pm0.01$ & $0.010\pm0.03$ & $0.030\pm0.02$ & & 0.03 \\  
age, Gyr         & 12.0           & 13.5           & \textbf{12.5}  & \textbf{12.5}  & 12.5           & & \\
distance, kpc    & \textbf{7.5}   & \textbf{7.4}   & 7.3            & 7.3            & \textbf{7.6}   & & \\
\hline
$E(F555W-F814W)$ & $0.073\pm0.06$ & $0.080\pm0.04$ & $0.066\pm0.03$ & $0.044\pm0.04$ & $0.066\pm0.04$ & & 0.04 \\
age, Gyr         & {\it 12.0}     & {\it 12.0}     & {\it 12.0}     & {\it 12.0}     & {\it 12.0}     & &  \\
distance, kpc    & {\it 7.4}      & {\it 7.4}      & {\it 7.4}      & {\it 7.4}      & {\it 7.4}      & & \\
\hline
$E(G_\mathrm{BP}-G_\mathrm{RP})$ & $0.080\pm0.02$  & $0.080\pm0.04$ & $0.093\pm0.01$   &   & $0.013\pm0.03$ & & 0.04 \\
age, Gyr                         & 11.5            & 12.5           & \textbf{11.5}    &   & 12.5           & & \\
distance, kpc                    & \textbf{7.6}    & \textbf{7.6}   & 7.4              &   & \textbf{8.4}   & & \\
\hline
$E(U-B)$         & $0.046\pm0.06$ & $0.076\pm0.06$ & $0.061\pm0.05$ & $-0.008\pm0.06$ & $0.019\pm0.06$ & & 0.02 \\
age, Gyr         & 11.5           & 12.5           & \textbf{12.0}  & \textbf{11.0}   & 12.5           & & \\
distance, kpc    & \textbf{7.6}   & \textbf{7.1}   & 7.0            & 7.8             & \textbf{7.6}   & & \\
\hline
$E(B-V)$         & $0.060\pm0.04$ & $0.080\pm0.07$ & $0.050\pm0.04$ & $0.050\pm0.05$ & $0.030\pm0.05$  & & 0.03 \\ 
age, Gyr         & 12.5           & 13.5           & \textbf{13.0}  & \textbf{12.0}  & 13.5            & & \\
distance, kpc    & \textbf{7.1}   & \textbf{7.0}   & 6.8            & 7.0            & \textbf{7.2}    & & \\
\hline
$E(V-R)$         & $-0.003\pm0.04$ & $0.016\pm0.04$ & $0.019\pm0.03$ & $0.013\pm0.04$ & $-0.009\pm0.06$ & & 0.02 \\
age, Gyr         & 11.5            & 12.5           & \textbf{11.5}  & \textbf{12.0}  & 12.0            & & \\
distance, kpc    & \textbf{7.9}    & \textbf{7.5}   & 7.3            & 7.4            & \textbf{8.1}    & & \\
\hline
$E(R-I)$         & $0.062\pm0.03$ & $0.062\pm0.04$ & $0.051\pm0.02$ & $0.038\pm0.04$ & $0.041\pm0.03$ & & 0.02 \\
age, Gyr         & 12.0           & 14.0           & \textbf{13.5}  & \textbf{13.5}  & 14.0           & & \\
distance, kpc    & \textbf{7.6}   & \textbf{7.3}   & 7.0            & 7.1            & \textbf{7.7}   & & \\
\hline
$E(u-g)$      & $0.125\pm0.06$ & $0.125\pm0.05$ & $0.114\pm0.05$ & $0.103\pm0.05$ & $0.070\pm0.06$ & $0.066\pm0.05$ & 0.03 \\
age, Gyr      & 11.5           & 11.5           & 11.5           & 11.5           & 11.5           & 12.6           & \\
distance, kpc & 7.0            & 7.0            & 6.7            & 6.9            & 7.2            & 7.4            & \\
\hline
$E(g-r)$      & $0.042\pm0.03$ & $0.048\pm0.04$ & $0.032\pm0.04$ & $0.021\pm0.04$ & $0.053\pm0.03$ & $0.058\pm0.04$ & 0.03 \\
age, Gyr      & 11.5           & 13.5           & 12.0           & 13.0           & 13.0           & 12.6           & \\
distance, kpc & 7.7            & 7.4            & 7.2            & 7.2            & 7.0            & 7.2            & \\
\hline
$E(r-i)$      & $0.053\pm0.02$ & $0.050\pm0.02$ & $0.056\pm0.02$ & $0.053\pm0.02$ & $0.062\pm0.02$ & $0.047\pm0.02$ & 0.02 \\
age, Gyr      & 11.5           & 13.0           & 11.5           & 12.5           & 13.0           & 12.6           & \\
distance, kpc & 7.2            & 7.2            & 6.8            & 6.7            & 6.7            & 7.2            & \\
\hline
$E(i-z)$      & $0.023\pm0.02$ & $0.026\pm0.02$ & $0.029\pm0.02$ & $0.026\pm0.02$ & $0.044\pm0.02$ & $0.012\pm0.02$ & 0.02 \\
age, Gyr      & 12.0           & 12.5           & 11.5           & 12.5           & 12.0           & 12.6           & \\
distance, kpc & 7.8            & 7.6            & 7.4            & 7.0            & 7.1            & 7.5            & \\
\hline
\end{tabular}
\]
\end{table*}


\begin{table*}
\def\baselinestretch{1}\normalsize\normalsize
\contcaption{
}
\label{fit:continued}
\[
\begin{tabular}{lccccccc}
\hline
       & PARSEC & MIST  & DSEP  & Old BaSTI & New BaSTI & APM & Harris \\
\hline
$E(g_\mathrm{P1}-r_\mathrm{P1})$  & $0.067\pm0.02$ & $0.077\pm0.03$ & $0.058\pm0.02$ & & & & 0.03 \\       
age, Gyr            & 12.5           & 14.0           & \textbf{13.0}  & & & & \\
distance, kpc       & \textbf{7.0}   & \textbf{6.8}   & 6.8            & & & & \\
\hline
$E(r_\mathrm{P1}-i_\mathrm{P1})$  & $0.024\pm0.01$ & $0.024\pm0.03$ & $0.021\pm0.01$ & & & & 0.02 \\
age, Gyr            & 12.0           & 14.0           & \textbf{13.0}  & & & & \\
distance, kpc       & \textbf{7.2}   & \textbf{7.2}   & 7.1            & & & & \\
\hline
$E(i_\mathrm{P1}-z_\mathrm{P1})$  & $0.019\pm0.01$ & $0.017\pm0.01$ & $0.020\pm0.01$ & & & & 0.015 \\
age, Gyr            & 11.5           & 14.0           & \textbf{12.0}  & & & & \\
distance, kpc       & \textbf{7.7}   & \textbf{7.2}   & 7.2            & & & & \\
\hline
$E(z_\mathrm{P1}-y_\mathrm{P1})$  & $0.008\pm0.01$ & $0.011\pm0.02$ & $0.011\pm0.01$ & & & & 0.008 \\
age, Gyr            & 12.0           & 13.5           & \textbf{13.0}  & & & & \\
distance, kpc       & \textbf{7.2}   & \textbf{7.0}   & 6.8            & & & & \\
\hline
$E(F814W-K_s)$  & $0.121\pm0.04$ & $0.113\pm0.04$ & $0.136\pm0.04$ & &   $0.098\pm0.04$ & & 0.04 \\   
age, Gyr        & 10.5           & 12.0           & \textbf{11.0}  & &   11.0           & & \\
distance, kpc   & \textbf{7.7}   & \textbf{7.5}   & 7.4            & &   \textbf{7.6}   & & \\
\hline
$E(G-H_\mathrm{UKIDSS})$ & $0.127\pm0.03$ & $0.095\pm0.04$ & $0.159\pm0.01$ & & & & 0.06 \\
age, Gyr                 & 11.5           & 13.0           & \textbf{12.0}  & & & & \\
distance, kpc            & \textbf{7.6}   & \textbf{7.6}   & 7.2            & & & & \\
\hline
$E(G-K_s)$    & $0.145\pm0.06$ & $0.134\pm0.06$ & $0.179\pm0.03$ & &  $0.246\pm0.10$ & & 0.07 \\
age, Gyr      & 11.5           & 13.0           & \textbf{12.5}  & & 12.0            & & \\
distance, kpc & \textbf{7.2}   & \textbf{7.0}   & 6.8            & & \textbf{7.0}    & & \\
\hline
$E(G-W1)$             & $0.213\pm0.04$   & $0.150\pm0.04$   & $0.213\pm0.04$  &   & $0.300\pm0.10$  & & 0.075 \\  
age, Gyr              & 11.5             & 13.0             & \textbf{12.0}   &   & 12.0            & & \\
distance, kpc         & \textbf{7.5}     & \textbf{7.6}     & 7.5             &   & \textbf{7.4}  & & \\
\hline
$E(I-H_\mathrm{UKIDSS})$ & $0.055\pm0.04$ & $0.061\pm0.03$ & $0.061\pm0.02$ & & & & 0.04 \\
age, Gyr                 & 11.0           & 12.5           & \textbf{11.5}  & & & &  \\
distance, kpc            & \textbf{7.6}   & \textbf{7.6}   & 7.4            & & & &  \\
\hline
$E(z-K_s)$    & $0.041\pm0.03$ & $0.046\pm0.03$ & $0.064\pm0.02$ & & $-0.023\pm0.03$ & & 0.03 \\
age, Gyr      & 11.5           & 13.0           & 12.0           & & 13.0            & & \\
distance, kpc & 7.5            & 7.4            & 7.0            & & 7.2             & &  \\
\hline
$E(H_\mathrm{UKIDSS}-W1)$ & $0.051\pm0.02$  & $0.047\pm0.02$ & $0.037\pm0.02$ & & & & 0.012 \\
age, Gyr                  & {\it 12.0}      & {\it 12.0}     & {\it 12.0}     & & & & \\
distance, kpc             & {\it 7.4}       & {\it 7.4}      & {\it 7.4}      & & & &  \\
\hline
$E(J-K_s)$    & $0.059\pm0.02$ & $0.037\pm0.04$ & $0.048\pm0.01$ & & $0.016\pm0.03$ & & 0.016 \\    
age, Gyr      & 11.5           & 12.5           & \textbf{12.0}  & & 13.0           & & \\
distance, kpc & \textbf{7.1}   & \textbf{7.2}   & 6.7            & & \textbf{7.4}   & & \\
\hline
$E(y_\mathrm{P1}-W1)$   & $0.092\pm0.05$ & $0.092\pm0.04$ & $0.104\pm0.05$ & & & & 0.035 \\
age, Gyr      & 11.0           & 12.0           & \textbf{12.0}  & & & & \\
distance, kpc & \textbf{7.4}   & \textbf{7.4}   & 7.0            & & & & \\
\hline
\end{tabular}
\]
\end{table*}


\section{Results}
\label{results}

Our results for different colours are consistent among themselves within their precision.
For example, the fitting of the PARSEC isochrones gives $E(F438W-F606W)=0.101$, $E(F606W-F814W)=0.045$,
$E(F814W-K_s)=0.121$ and $E(F438W-K_s)=0.267$ with $E(F438W-K_s)\approx E(F438W-F606W)+E(F606W-F814W)+E(F814W-K_s)$,
as expected.
Another example is seen in Table~\ref{fit}: $E(G-H_\mathrm{UKIDSS})+E(H_\mathrm{UKIDSS}-W1)\approx E(G-W1)$
within their uncertainties.
Therefore, to avoid redundancy in this paper, we have chosen to only show the fits for some selected key
colours rather than a full possible set of more than 100 colours.
Fig.~\ref{best336_438} -- \ref{bestzks} 
in the Appendix show the CMDs with the best isochrone fits for the key colours.
All other CMDs can be provided on request.
The age, distance, and reddening derived from some isochrone fits are presented in Table~\ref{fit}
together with the same reddenings calculated from the Harris's $E(B-V)=0.03$ mag and
the \citetalias{ccm89} extinction law with $R_\mathrm{V}=3.1$.

The maximum offset of the isochrone from the fiducial sequence along the reddening direction in the CMD
within the RGB, SGB, and a brighter part of the MS is accepted as an accuracy of the determination of the reddening
and shown in Table~\ref{fit} after the values of the reddening.
This uncertainty dominates over any random or systematic uncertainty of the photometry.

In the UV there are few cases when the isochrone cannot fit the fiducial sequence better than to 0.1 mag.
These cases are seen in Table~\ref{fit} and Fig.~\ref{best336_438}: PARSEC almost fails for
$F275W-F336W$ and $F336W-F438W$ colours, while DSEP fails as well for $F275W-F336W$ colour.
This is a well-known imperfection of the models in the UV \citepalias{barker2018}, as mentioned in Sect.~\ref{intro}.
Although we present these cases in Table~\ref{fit} and in the figures, we do not use them for calculating
the average distance and age.

Also we do not take into account the offsets of all the isochrones from the data at the lower MS in
Fig.~\ref{best555_814}.
This is the well-known poor representation of the physics of red dwarfs by all the models used \citep{layden2005}.

Some our CMDs can be compared with those presented by \citet{casagrande2014, casagrande2018a}.
They have compared the same {\it HST}/ACS, $UBVRI$, Pan-STARRS, and SDSS photometry of NGC\,5904 with a set of the 
Victoria-Regina isochrones.
They adopted the reddenings and extinction law from \citet{schlaflyfinkbeiner2011}.
In particular, our Fig.~\ref{best606_814} can be compared with the figure 10 from \citet{casagrande2014},
while our Fig.~\ref{bestpanstarrs} with the figure 7 from \citet{casagrande2018a}.
Usually their isochrones lie about 0.05 mag to the red of the fiducial sequences.
In some cases this offset is different for the RGB and MS in the same CMD.
To achieve a fully consistent fit (to make the isochrones bluer) one has to adopt an improbable
nearly zero reddening, much lower than the adopted estimate $E(B-V)=0.032$ mag from 
\citet{schlaflyfinkbeiner2011}.
Thus, different models/isochrones may be more or less reliable in such a fitting.
Consequently, it is clearly advantageous to use different models/isochrones to fit CMDs.
Also, it is evident from our results that PARSEC, MIST, DSEP, and BaSTI are quite reliable in this task.

The distance to NGC\,5904 is determined better for the models and
data for both the HB and TO (PARSEC, MIST, and new BaSTI, which obsoletes old BaSTI).
However, we do not use the PARSEC $F275W-F336W$ and $F336W-F438W$ colours with the failed fitting,
all the $UBVRI$-{\it WISE} colours due to their noisy TO,
$H_\mathrm{UKIDSS}-W1$ colour due to its poor separation of the HB and TO from the MS,
\footnote{All the stars have nearly the same $H_\mathrm{UKIDSS}-W1$ colour.
In this case we adopt a distance of 7.4 kpc and highlight it in Table~\ref{fit} with italic.}
and all SDSS colours due to their large systematic offsets discussed below.
The values used to derive the distance are
highlighted in Table~\ref{fit} with bold.
By use of them, we calculate the most probable distance to NGC\,5904
as $7386\pm311$~pc, i.e. a true distance modulus of $(m-M)_0=14.34\pm0.09$ mag.
The apparent $V$-band distance modulus based on this estimate is compared with the literature in Sect.~\ref{extlaw}.

From our estimate of the distance we calculate the parallax $0.135\pm0.006$ mas.
This agrees with recent estimate of the parallax from the {\it Gaia} DR2 $0.113+0.029=0.142\pm0.040$ mas,
where $0.029$ and $0.040$ mas are the parallax global zero-point correction 
derived from the analysis of the high-precision quasar sample 
and the median uncertainty in parallax, respectively \citep{lindegren2018}.

The age might be determined better for models with both the solar-scaled and He-$\alpha$-enhanced abundances
(DSEP and old BaSTI) fitted to the
data on the SGB.
However, we do not use the DSEP $F275W-F336W$ colour with the failed fitting,
all the $UBVRI$-{\it WISE} colours,
the $H_\mathrm{UKIDSS}-W1$ colour (its adopted age of 12 Gyr is highlighted in Table~\ref{fit} with italic),
and all the SDSS colours for the same reasons as for the distance.
The used values are highlighted in Table~\ref{fit} with bold.
They provide the age $12.15\pm0.70$ Gyr.
This is in line with the
age and its uncertainty derived
from \citetalias{an2009}: either 11.2 or 12.6 Gyr.
The fitting is almost equally good for both the ages. Therefore, the age can be determined
from \citetalias{an2009} as $11.9\pm0.7$ Gyr.
Yet, we have found in our fitting that the age for NGC\,5904 is
quite uncertain: its variation by $\pm1$ Gyr from the best fit gives an almost equally good fitting.
Thus, our final estimate of the age is $12.15\pm1.00$ Gyr.
This value agrees well with different estimates from the literature.
Few cases outside this range in Table~\ref{fit} can be explained by an imperfection of the models.

The obtained distance and age values show no dependence on colour.
This is not the case for the obtained values of reddening. They must represent a dependence of extinction on
wavelength, i.e. an extinction law.
Each model and each series of the data gives its own set of extinctions and its own extinction law.
Below, we compare them with each other and with the law of \citetalias{ccm89}.

\begin{figure*}
\includegraphics{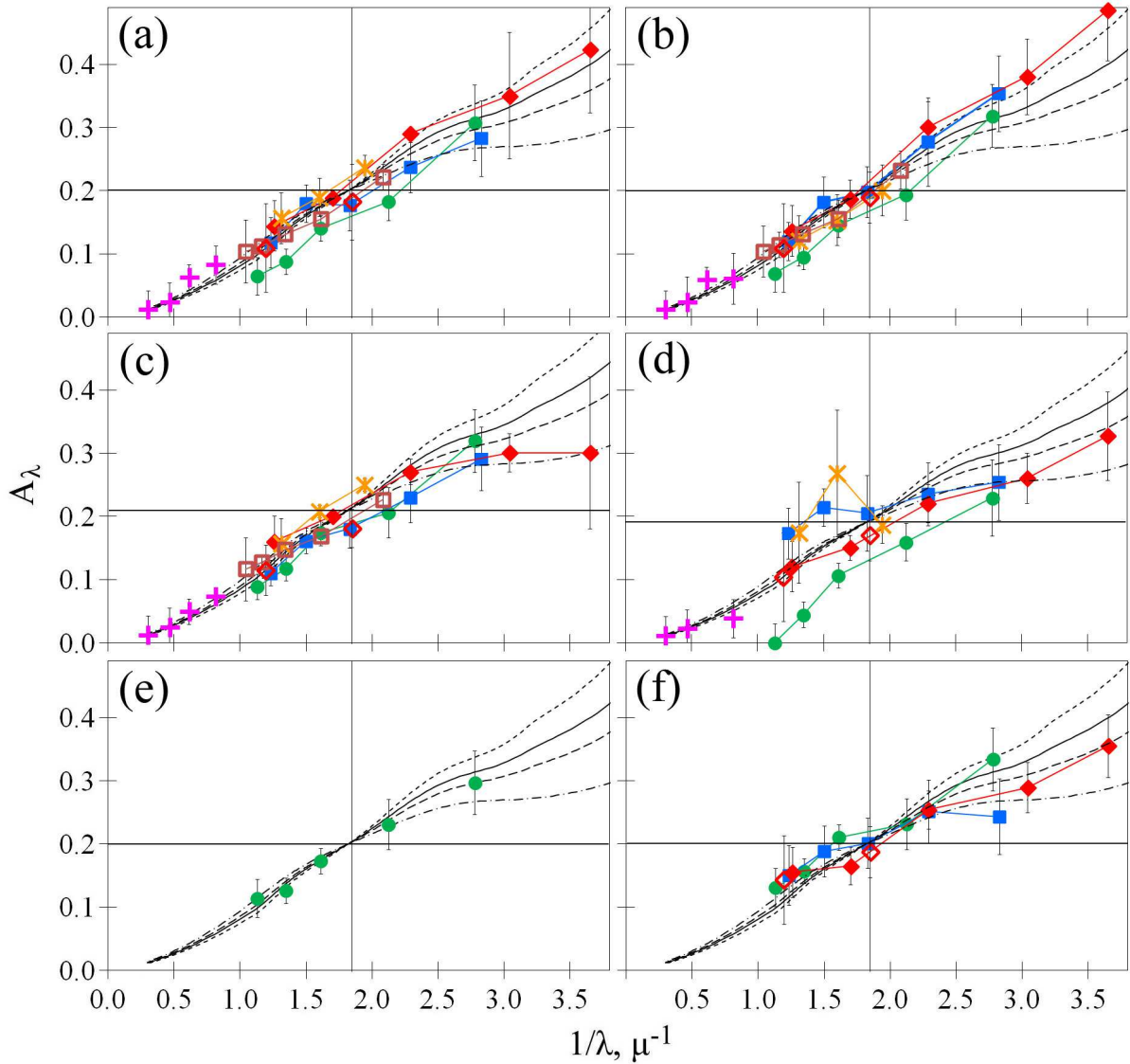}
\caption{The extinction laws from the isochrone fitting by (a) PARSEC, (b) MIST, (c) DSEP, (d) new BaSTI,
(e) \citetalias{an2009}, and (f) old BaSTI.
The {\it HST} series -- red diamonds (open ones for the WFPC2 data),
{\it Gaia} series -- yellow snowflakes,
$UBVRI$ series -- blue squares,
SDSS series -- green circles,
Pan-STARRS series -- brown squares, and
IR series -- purple crosses.
The black dotted, solid, dashed and dash-dotted black curves show the \citetalias{ccm89} extinction law with
$R_\mathrm{V}=2.6$, 3.1, 3.6, and 5, respectively.
}
\label{law}
\end{figure*}

\subsection{Extinction law}
\label{extlaw}

PARSEC, MIST, DSEP and new BaSTI, in contrast to APM and old BaSTI, provide us with IR colours.
The IR extinction and its variation due to some reasonable variations of the extinction law are very low.
For example, $E(B-V)=0.03$ with the \citetalias{ccm89} extinction law with $R_\mathrm{V}=3.1$ corresponds to
$A_\mathrm{K_s}=0.012$ and $A_\mathrm{W1}=0.006$ mag.
Anyway, such values are lower than a typical systematic uncertainty of 0.02 mag for the fitting of the CMDs with 
the $K_s$ and $W1$ colours (see Table~\ref{fit}).
Hence, our calculations of the optical and UV extinctions are based on the fact that the IR extinctions used
are nearly zero and almost independent of the extinction law.

Namely, given $A_\mathrm{K_s}=0.012$ and $A_\mathrm{W1}=0.006$ mag, we calculate the extinctions for the 
remaining 27 bands,
including $A_\mathrm{B}$ and $A_\mathrm{V}$, then calculate $R_\mathrm{V}=A_\mathrm{V}/(A_\mathrm{B}-A_\mathrm{V})$
and use the \citetalias{ccm89} extinction law with this $R_\mathrm{V}$ 
to re-calculate $A_\mathrm{K_s}$ and $A_\mathrm{W1}$.
The iterations of this procedure converge to some extinction values,
including $A_\mathrm{K_s}=0.02$ and $A_\mathrm{W1}=0.01$,
\footnote{The difference between $E(G-K_s)$ and $E(G-W1)$ in Table~\ref{fit} provides $E(K_s-W1)=0.04\pm0.03$
as the average value from PARSEC, MIST, DSEP, and new BaSTI. This does not contradict $E(K_s-W1)=0.01$ in this solution.}
as well as
$A_\mathrm{V}=0.20$, 0.20, 0.21 and 0.19 mag for PARSEC, MIST, DSEP, and new BaSTI, respectively,
by use of the {\it HST}, {\it Gaia}, $UBVRI$ and Pan-STARRS data series.
These values are shown in Fig.~\ref{law} by the black horizontal lines, while the $V$ band -- by the black vertical lines.
This Figure shows all the derived extinctions as some functions of $1/\lambda$ (i.e. as extinction laws) 
for the fits by
(a) PARSEC, (b) MIST, (c) DSEP, and (d) new BaSTI.
The different series of data are shown by different colours.
The uncertainties of the extinctions are calculated from the uncertainties of the reddenings and
shown in Fig.~\ref{law} by the vertical bars.
The black curves show the \citetalias{ccm89} extinction law with different $R_\mathrm{V}$.

The optical-IR and IR-IR CMDs and the reddenings (isochrone-fiducial colour offsets) derived from them
appear the key data in our study. Hence, we note that the extinctions for the five optical data series are based on
the derived
(i) $E(F814W-K_s)$, 
(ii) $E(G-W1)$ and $E(G_\mathrm{RP}-W1)$, 
(iii) $E(I-H_\mathrm{UKIDSS})$ and $E(H_\mathrm{UKIDSS}-W1)$ [$E(I-W1)$ for new BaSTI], 
(iv) $E(y_\mathrm{P1}-W1)$, and
(v) $E(z-K_s)$
for the {\it HST}, {\it Gaia}, $UBVRI$, Pan-STARRS, and SDSS series, respectively.
Thus, all these extinctions from the five data series by the four models are based on the adopted 
$A_\mathrm{K_s}$ or $A_\mathrm{W1}$.

Fig.~\ref{law} shows a good agreement between the extinction laws from the different models,
except the UV bands, as well as between the different data series,
except a noticeable negative offset of the SDSS series for all the models.
The SDSS photometry may be less accurate than the one from {\it HST}, {\it Gaia}, $UBVRI$ and Pan-STARRS
(see, for example, a discussion on some errors of the SDSS photometry by \citealt{an2008, dotter2008}).
Therefore, the SDSS results are not used for calculating the final average values of the extinctions, distance and age.

The average $A_\mathrm{V}$ is $0.2$ mag.
This is the main result of our study.
Its systematic uncertainty is dominated by a systematic uncertainty of about $0.02$ mag
of the zero-point IR extinctions $0<A_\mathrm{K_s}<0.04$ and $0<A_\mathrm{W1}<0.02$.
These constraints include all possible variations of these extinctions due to any variation of any real or
adopted extinction law.
Therefore, in fact, $A_\mathrm{K_s}$, $A_\mathrm{W1}$ and all other extinctions, derived by use of 
PARSEC, MIST, DSEP and new BaSTI, are not tied to the \citetalias{ccm89} or any other extinction law.
Therefore they can be used to analyse a real extinction law in the direction of NGC\,5904.
Fig.~\ref{law} shows that all the extinction values, except $A_\mathrm{F275W}$ for MIST and some for the SDSS,
agree with the \citetalias{ccm89} extinction law with $R_\mathrm{V}=3.60\pm0.05$ (dashed curve) within their precision.
Thus, we conclude that the data for NGC\,5904 do not need an unusual extinction law.

With the estimate $A_\mathrm{V}=0.2$ mag we obtain an apparent $V$-band distance modulus of
$(m-M)_\mathrm{V}=14.54\pm0.11$ mag, which includes the additional error inherent in $A_\mathrm{V}$.
This agrees with $(m-M)_\mathrm{V}=14.46$ from
\citet{harris} and $14.47\pm0.07$ from \citet{vandenberg2018} obtained as an average of various estimates.

In contrast to PARSEC, MIST, DSEP and new BaSTI, \citetalias{an2009} and old BaSTI have no IR colour.
Hence, their extinctions have to be based on extinction estimates for their reddest bands from
\citetalias{ccm89} extinction law with $R_\mathrm{V}$ as a free parameter.
All their extinction series are reduced to $A_\mathrm{V}=0.2$, on average.
We derive $R_\mathrm{V}=5$ and, hence, $E(B-V)=0.2/5=0.04$ mag for both APM and old BaSTI.
Fig.~\ref{law} (e), (f) shows the extinctions derived by use of APM and old BaSTI, respectively.
Yet, these incomplete results only show that their reddenings in the optical and UV ranges agree with those from
PARSEC, MIST, DSEP and new BaSTI, when a zero-point extinction is fixed.

We have found a rather high optical extinction for NGC\,5904.
It is about twice higher than the `canonical' $A_\mathrm{V}=0.09$ mag
(as a product of $E(B-V)=0.03$ by $R_\mathrm{V}=3.1$) from \citet{harris}.
This may be related to
the well-known issue that a `larger reddening than suggested by the adopted reddening curve is needed
in some cases [for open and globular clusters]' \citep{dotter2008}.

As evident from Fig.~\ref{law} and Table~\ref{fit}, this high extinction is due to rather high reddenings
between the IR and the optical bands.
It is true as for long (such as $G-W1$), as for short (such as $H_\mathrm{UKIDSS}-W1$) colour baselines.
A lower part of Table~\ref{fit} with the optical-IR and IR-IR colours shows that the derived reddenings tend to be twice
as high as those calculated from Harris's $E(B-V)=0.03$ mag.
Taking into account nearly zero IR extinctions, this means a rather high extinction in the range $625<\lambda<2000$ nm.
Indeed, it is seen from Fig.~\ref{law} that the slopes of the data series within this range (i.e. $0.5<1/\lambda<1.6$)
tend to be steeper than within $1.6<1/\lambda<2.0$, near the $V$ band.
Thus, the range $625<\lambda<2000$ nm produces the bulk of the value $A_\mathrm{V}=0.2$.

\begin{figure*}
\includegraphics{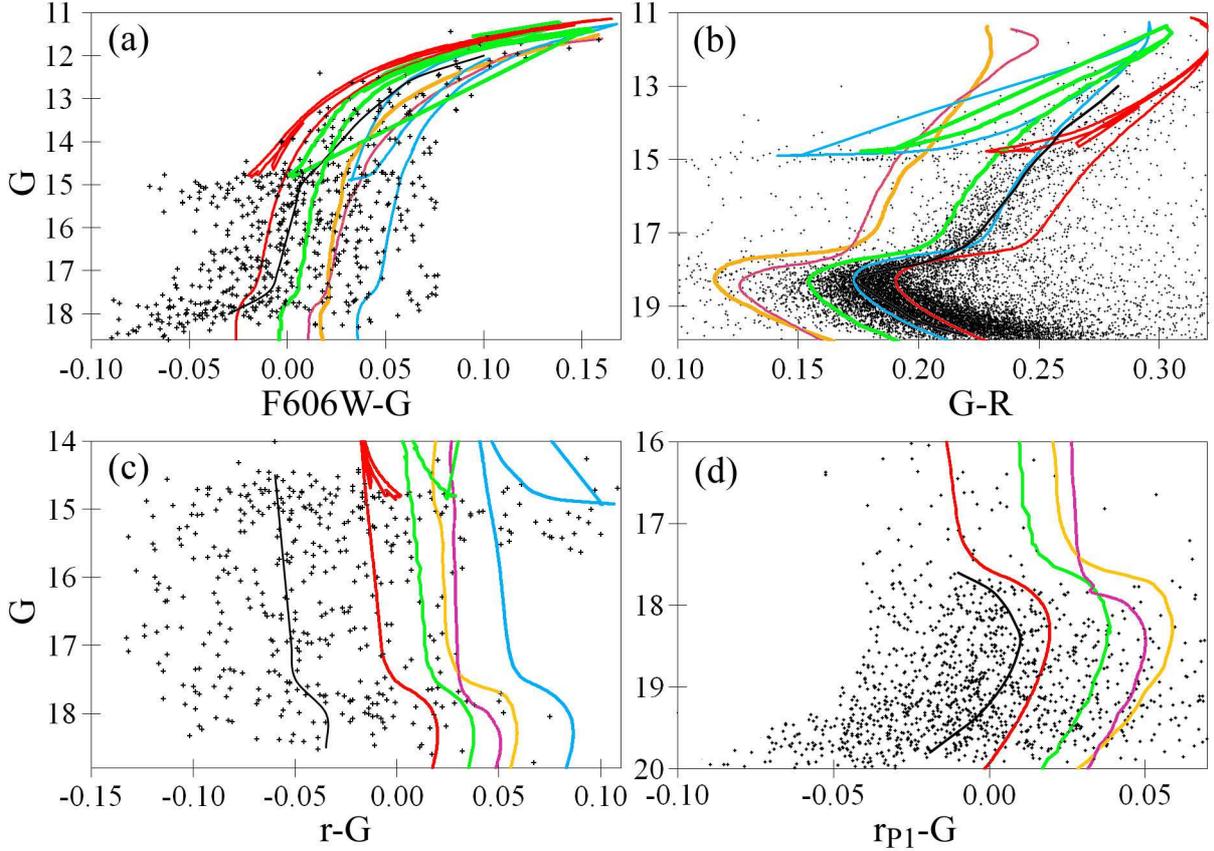}
\caption{The CMDs of NGC\,5904
(a) $F606W-G$ versus $G$,
(b) $G-R$ versus $G$,
(c) $r-G$ versus $G$, and
(d) $r_\mathrm{P1}-G$ versus $G$
with the fiducial sequences (black curve) and the isochrones for the age 12 Gyr, distance 7.4 kpc, reddening
$E(B-V)=0.03$~mag and \citetalias{ccm89} extinction law with $R_V=3.1$, coloured as in Fig.~\ref{f814_ks}.
}
\label{offset}
\end{figure*}

\begin{figure*}
\includegraphics{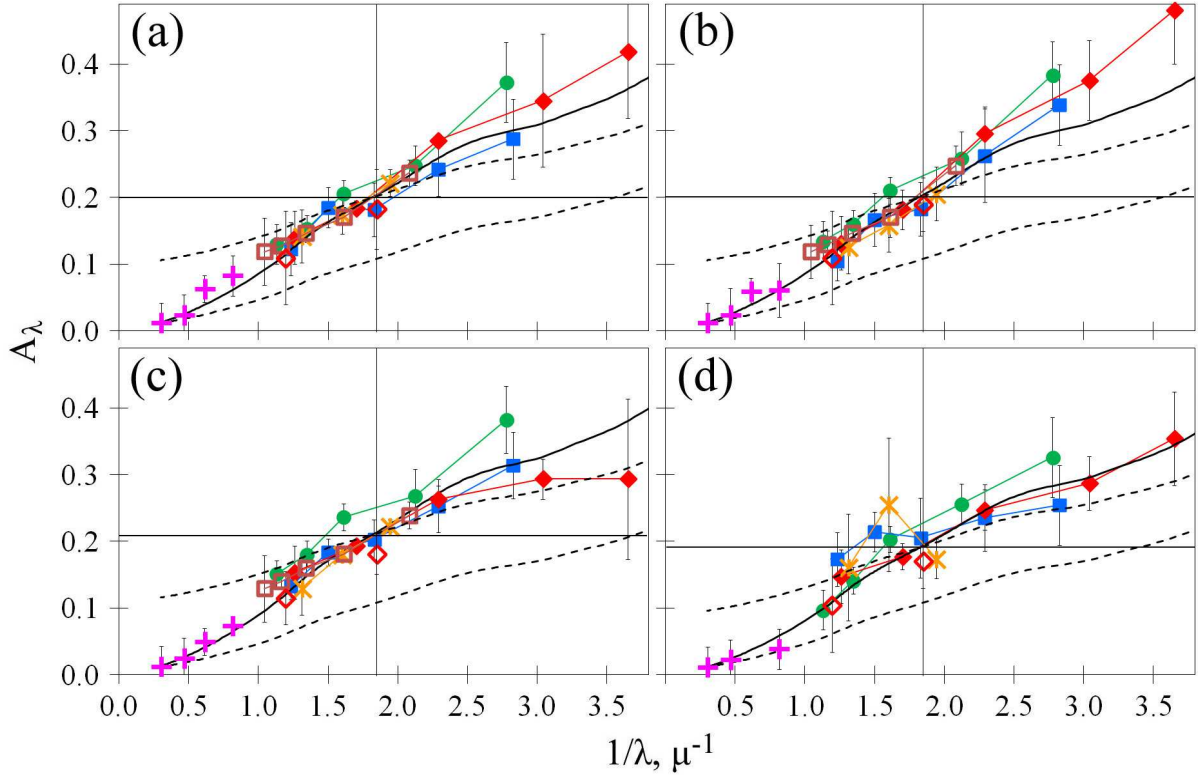}
\caption{The extinction laws from the isochrone fitting by (a) PARSEC, (b) MIST, (c) DSEP, and (d) new BaSTI
after the adjustment for the systematic offsets from Table~\ref{tableoffset}.
The designation of the data series as in Fig.~\ref{law}.
The black solid curve shows the \citetalias{ccm89} extinction law with $A_\mathrm{V}$ derived for the models and
$R_\mathrm{V}=3.6$.
The black lower dotted curve shows the \citetalias{ccm89} extinction law with $A_\mathrm{V}=0.09$ mag and
$R_\mathrm{V}=3.1$, while the upper one shows the same but shifted up to $A_\mathrm{V}$ derived for the models.
}
\label{adjust}
\end{figure*}

\section{Discussion}
\label{discuss}

The large offsets of the isochrones from the observations in the optical-IR and IR-IR CMDs may be explained by
some inherent offsets of the model colours by about 0.1 mag, similar for all models, as well as for all colours,
which are close to each other.

Moreover, such inherent offsets of the model colours, being different for different data series, may explain
the vertical offsets (i.e. the different extinction laws) of the data series, which are seen in Fig.~\ref{law}.

To test these assumptions, we analyse some CMDs with rather close colour bands, which have a difference of their 
extinction coefficients $<0.2$.
In this case, any reasonable variation of the extinction and reddening (e.g. within $0.02<E(B-V)<0.12$ mag) 
does not shift any isochrone in this CMD along the colour by more than $0.2\times(0.12-0.02)=0.02$ mag.
This is negligible w.r.t. the colour offsets under consideration.
It appears that any reasonable variation of [Fe/H], abundance, age, and distance also does not shift these isochrones
by more than 0.02 mag.
This means that any considerable colour offset between an isochrone and the fiducial sequence in such a CMD is due to
an imperfection of this isochrone.

A quite interesting range, $588<\lambda_\mathrm{eff}<667$ nm, contains 5 close bands:
$F606W$, $G$, $r$, $R$ and $r_\mathrm{P1}$ (see Table~\ref{filters}).
We consider only their colours with the extinction coefficients $<0.2$.
Four CMDs with such colours, together with the isochrones from PARSEC, MIST, DSEP, and new BaSTI for the
age 12 Gyr,
distance 7.4 kpc, $E(B-V)=0.03$ and \citetalias{ccm89} extinction law with $R_V=3.1$ are presented in Fig.~\ref{offset}.
All the CMDs show offsets of the isochrones from the fiducial sequences (black curves) within $\pm0.05$ mag,
except those with the SDSS $r$ band with some offsets up to 0.11 mag.
These offsets are different for different models/isochrones in the same CMD, as well as for different series/colours.
This is explained, among other reasons, by quite different colour-$T_\mathrm{eff}$ relations and bolometric corrections 
in these models, as discussed by \citet{paxton2011, choi2016, newbasti}.
Therefore, Fig.~\ref{offset} shows that the situation when the four models (PARSEC, MIST, DSEP and new BaSTI)
provide by chance the same inherent optical-IR colour offsets of about 0.1 mag for the four data series 
({\it HST}, {\it Gaia}, Pan-STARRS and $UBVRI$) is quite unlikely. 
It is even more unlikely that some large inherent colour offsets for the bands, at least, from $F438W$ to $K_s$, 
i.e. within $430<\lambda<2200$ nm, precisely follow by chance the \citetalias{ccm89} extinction law
[i.e. provide the small scatter of the symbols around a black curve in Fig.~\ref{law}~(a)--(d)].
Hence, the discovered optical-IR colour offsets seem to be not inherent, but generated by reddening.

Moreover, the comparison of the MIST, PARSEC, DSEP and BaSTI isochrones in the plane of luminocity versus 
T$_\mathrm{eff}$ for $Z=0.0001$ and age 10 Gyr, i.e. for typical parameters for GCs, by \citet{choi2016}
(see their figure 16) shows that for the same luminocity a difference in T$_\mathrm{eff}$ is always within $\pm150$ K.
This corresponds to $\pm0.03$~mag of a colour offset [see \citet{vandenberg2018}].
This is much less than the colour offset of about 0.1 mag, which needs to be explained.
Figure 18 of \citet{choi2016} confirms this: for an age of 10 Gyr any difference between the 
MIST, PARSEC and BaSTI optical-IR and IR-IR colours (even for $V-K$) is much less than 0.1 mag.
However, \citet{kucinskas2006}, \citet{hendricks2012} have found that an uncertainty of T$_\mathrm{eff}$ of 100 K can lead to
an uncertainty of the optical-IR colours of the RGB and AGB stars up to 0.05 mag.
This is another reason why we have paid more attention to the TO, SGB, MS than to the RGB and AGB in our isochrone 
fitting.
Finally, we are inclined to conclude with a caution that the large derived colour offsets between the isochrones and 
the fiducial
sequences in the optical-IR and IR-IR CMDs are due to extinction rather than to inherent model colour offsets.

\begin{table}
\def\baselinestretch{1}\normalsize\normalsize
\caption[]{The offset correrctions (mag) applied for the various data series and models.
}
\label{tableoffset}
\[
\begin{tabular}{lrrrr}
\hline
\noalign{\smallskip}
 & \multicolumn{4}{c}{Model} \\
\noalign{\smallskip}
 Series & PARSEC & MIST & DSEP & new BaSTI \\
\hline
\noalign{\smallskip}
{\it HST}  & $-0.005$ & $-0.005$ & $-0.007$ & 0.026     \\
{\it Gaia} & $-0.015$ & 0.005    & $-0.027$ & $-0.013$  \\
$UBVRI$    & 0.005    &	$-0.015$ & 0.022    & $-0.013$  \\
Pan-STARRS  & 0.015    &	0.015    & 0.012    &           \\
SDSS       & 0.065    & 0.065    & 0.062    & 0.096 \\
\hline
\end{tabular}
\]
\end{table}


To explain the difference between the extinction laws for the different data series in Fig.~\ref{law},
we adjust the colour offsets from Fig.~\ref{offset} assuming the means of the
{\it HST}, {\it Gaia}, $UBVRI$ and Pan-STARRS extinctions as true and, hence, fixing them.
This gives us some constant corrections to the extinction laws for the various data series and models,
which are presented in Table~\ref{tableoffset}.
These corrections, except for SDSS, are nearly zero.
This means that the inter-series CMDs (such as in Fig.~\ref{offset}) and their reddenings are in line with the 
intra-series CMDs (as in Fig.~\ref{f814_ks}--\ref{gw1} and \ref{best336_438}--\ref{bestzks}) and their reddenings.
This, in turn, means that the series agree with each other.

The extinction laws from PARSEC, MIST, DSEP, and new BaSTI with these corrections applied are shown in
Fig.~\ref{adjust}, which is similar to Fig.~\ref{law}.
This adjustment decreases the scatter of the data series, except the SDSS.

The black solid curve in Fig.~\ref{adjust} shows the \citetalias{ccm89} extinction law with $R_\mathrm{V}=3.6$
and with $A_\mathrm{V}$ derived for the models.
As in Fig.~\ref{law}, it is seen that all the extinctions, except $A_\mathrm{F275W}$ for MIST and some for the SDSS,
agree with this law within their precision.

The black dotted curves in Fig.~\ref{adjust} show the \citetalias{ccm89} extinction law with
$R_\mathrm{V}=3.1$ for $A_\mathrm{V}=0.09$ mag (lower curve) and $A_\mathrm{V}$ derived for the models (upper curve).
Thus, the lower curve extrapolates the Harris's value of $E(B-V)=0.03$ mag out of the range between the $B$ and $V$ bands.
The upper curve is drawn to emphasise that all the data series, including also the {\it HST} WFPC2
(the red open diamonds),
being scarcely agreed with $E(B-V)=0.03$ mag (the slope of the upper dotted curve),
yet, demonstrate a steeper slope (i.e. $E(B-V)>0.03$ mag), which better agrees with the slope of the solid curve
(i.e. $E(B-V)=0.055$ mag).
This is a very important result, since the reddenings derived in our fits (i.e. the slopes of the series in the 
Fig.~\ref{law} and \ref{adjust})
are independent of any zero-point extinction.
In other words, these steep slopes and a positive minimal extinction are the only reasons for such a high $A_\mathrm{V}$.

The reddenings for the long-baseline colours [such as $E(F438W-W1)$] obtained by us show a good agreement with the
\citetalias{ccm89} extinction law with $R_\mathrm{V}=3.6$ and $A_\mathrm{V}=0.2$ mag.
However, some reddenings for the short-baseline colours from Table~\ref{fit} (such as $E(H_\mathrm{UKIDSS}-W1)$)
show some noticeable small-scale deviations from this law.
These deviations may explain the diversity of the reddening estimates discussed in Sect.~\ref{intro}.
On the one hand,
Table~\ref{fit} gives us the average $E(B-V)=0.054\pm0.020$ mag in a good agreement with the \citetalias{ccm89} extinction
law giving $E(B-V)=A_\mathrm{V}/R_\mathrm{V}=0.2/3.6=0.056$ mag.
Given its uncertainty of $\pm0.02$ mag, this estimate is in the best agreement with the one from \citet{green2018},
in a moderate agreement with those of \citet{drimmel} and \citetalias{sfd},
and in a poor agreement with the remaining reddening estimates discussed in Sect.~\ref{intro}, including
the Harris's $E(B-V)=0.03$ mag.
On the other hand,
Table~\ref{fit} gives us the averages $E(V-R)=0.007\pm0.02$ and $E(H_\mathrm{UKIDSS}-W1)=0.045\pm0.01$ mag.
With the extinction coefficients $E(V-R)/E(B-V)=0.67$ and $E(H_\mathrm{UKIDSS}-W1)/E(B-V)=0.468$ from the \citetalias{ccm89}
extinction law with $R_\mathrm{V}=3.6$, this gives us $E(B-V)=0.01\pm0.03$ and $0.10\pm0.02$ mag, respectively.
These values are completely inconsistent with each other.
The former and the latter agree best with the lowest reddening estimates from \citet{green2015} and \citet{arenou}
and the highest ones from \citet{av} and \citet{g17}, respectively.

Thus, as assumed in Sect.~\ref{intro}, some deviations of the real extinction law from
\citetalias{ccm89} may explain the diversity of the previous reddening estimates for NGC\,5904, which are
inter- or extrapolated by use of a `standard' extinction law from some observations at very different wavelengths.

Fortunately, we have found a real extinction law for NGC\,5904 rather close to the \citetalias{ccm89} with
a reasonable value of $R_\mathrm{V}$. However, it may not be the case for other GCs, which we intend to consider
in our future studies.

Our findings are in line with the earlier criticism of the \citetalias{ccm89} extinction law with $R_\mathrm{V}=3.1$
as the `standard' and universal one.
Some disadvantages of the \citetalias{ccm89} extinction law and the procedure of its creation are well-known
\citep{maiz2014}.
For example, \citet{fm2007} have found that there is no strong correlation between the UV and IR Galactic 
extinctions, in opposition to what \citetalias{ccm89} found.
Some different extinction laws with a higher IR extinction, with $R_\mathrm{V}>3.1$,
and with a grey extinction are typical far from the Galactic mid-plane
(NGC\,5904 is about 5.5 kpc above the mid-plane) instead of the \citetalias{ccm89} extinction law
\citep{gorbikov, rv, hendricks2012, g2013, davenport, g2016, astroph}.
Moreover, \citep{hendricks2012} have found $R_\mathrm{V}=3.62\pm0.07$ in the line of sight to the GC M4
as a result of a study, similar to ours. An excellent agreement of this result with our $R_\mathrm{V}=3.60\pm0.05$
for NGC\,5904 (M5) may be explained.
M4 and NGC\,5904 are in the same sky area affected by a higher extinction and unusual extinction law at
the upper part of the Gould Belt (near $l\approx15\degr$, $b\approx+19\degr$) \citep{gould, rv, astroph}.
`We find that dust at high latitude is neither quantitatively nor qualitatively consistent with standard reddening laws.'
\citep{peek2013}.

\section{Conclusions}
\label{conclusions}

In this study we have tested the expectation that an accurate fitting of the isochrones to a multi-band photometry
for a GC can provide some convergent estimates of its distance and age, together with a set of
estimates of the interstellar extinction in dependence on wavelength.
This set represents the real extinction law to the cluster.
We used the photometry of NGC\,5904 (M5) in 29 bands from
the {\it HST}, {\it WISE}, {\it Gaia} DR2, SDSS, UKIDSS, NTT, TNG, Pan-STARRS,
and a compilation of the $UBVRI$ photometry.
These bands cover a wavelength range from about 235 to 4070\,nm, i.e. from the UV to mid-IR.
To fit the data, we used the following five theoretical models of the stellar evolution:
PARSEC, MIST, DSEP, BaSTI (the old and new version), and the model from \citet{an2009}.
Some of the isochrones, which they produce, are calculated for non-scaled-solar abundances, with enhanced He
and $\alpha$-elements.
Yet, the fitting revealed that for NGC\,5904 the difference between scaled-solar and non-scaled-solar isochrones is
quite small, and both kinds of isochrones can be used.
We accept a metallicity of about [Fe/H]$=-1.33$ based on the spectroscopy taken from the literature.

We found for NGC\,5904 the age to be $12.15\pm1$ Gyr, distance $7386\pm311$\,pc,
true distance modulus $(m-M)_0=14.34\pm0.09$ and apparent $V$-band distance modulus $(m-M)_\mathrm{V}=14.54\pm0.11$.
These estimates agree with the literature estimates including the recent one from {\it Gaia} DR2 
$0.113+0.029=0.142$ mas, where $+0.029$ mas is the well-established global zero-point correction \citep{lindegren2018}.

From the fitting we found some systematic colour offsets of the isochrones from the fiducial sequences in
the CMDs. These offsets draw some monotonic functions of colour. Hence, they can be interpretted as sets of the
extinctions drawing an extinction law for each data series.
Nearly zero extinctions in $K_s$ and $W1$ bands, which are almost independent of extinction law, are used
as zero-point extinctions in the {\it HST}, {\it Gaia}, Pan-STARRS, $UBVRI$, and SDSS data series in their fitting by
PARSEC, MIST, DSEP, and new BaSTI.
As the main result of our study, all the extinction laws derived in the fitting of these data series, except the SDSS,
by these four models agree with each other and with the \citetalias{ccm89} extinction law with $R_\mathrm{V}=3.6$ and
$A_\mathrm{V}=0.20\pm0.02$ mag.
This uncertainty includes the uncertainties of the zero-point IR extinctions, of the extinction law, as well as
the scatter of the results of the fittings of the four data series by the four models.
Old BaSTI and the model from \citet{an2009}, providing no IR colour, cannot draw a complete extinction law.
Yet, their reddenings in the optical range do not contradict the remaining models.

All the data and models show agreed deviations from the \citetalias{ccm89} extinction law with $R_\mathrm{V}=3.6$
for some short-baseline colours, such as $V-R$ and $H_\mathrm{UKIDSS}-W1$.
These deviations can explain the diversity of the previous reddening estimates for NGC\,5904, which were
inter- or extrapolated by use of a `standard' extinction law from some observations at very different wavelengths.

The obtained $E(B-V)=0.054\pm0.020$ is in a moderate agreement with the generally accepted $E(B-V)=0.03$
from \citet{harris} and with the set of non-independent estimates of $E(B-V)\approx0.35$ from
\citet{2015ApJ...798...88M}, \citet{schlaflyfinkbeiner2011}, \citet{drimmel}, \citet{lallement2018}, and \citetalias{sfd}
within their uncertainties. 
However, due to the rather high $R_\mathrm{V}=3.6$, the obtained $A_\mathrm{V}=0.2$ mag is nearly twice as high as
the generally accepted $A_\mathrm{V}=0.09$ mag, as the product of $E(B-V)=0.03$ by $R_\mathrm{V}=3.1$.
A bulk of the extinction $A_\mathrm{V}=0.2$ mag appears within $625<\lambda<2000$ nm.
Hence, the extinction in this range cannot be reconciled with $R_\mathrm{V}=3.1$.

However, the obtained large colour offsets between the isochrones and the fiducial sequences in the CMDs
may also include some inherent offsets of model colours, e.g. due to some errors of
the colour-$T_\mathrm{eff}$ calibrations used.
Yet, such a large inherent colour offset of about 0.1 mag, similar in the fitting of the four independent data series
by the four rather different models seems to be rather unlikely.
Thus, we are inclined to conclude that a real extinction law for the dust medium between us and NGC\,5904 is close
to the \citetalias{ccm89} extinction law with $R_\mathrm{V}=3.6$ and $A_\mathrm{V}=0.2$ mag.
Nevertheless, a further improvement of the theoretical models and isochrones is highly needed.

\section*{Acknowledgements}

We thank an anonymous reviewer for useful comments.
We thank Santi Cassisi for providing the valuable BaSTI isochrones for {\it WISE},
Massimo Dall'Ora for providing the valuable $J$ and $K_s$ photometry for NGC\,5904,
Amina Helmi for providing the valuable {\it Gaia} DR2 data for NGC\,5904.
We thank Jieun Choi, Pavel Denissenkov, Aaron Dotter, Valery Kravtsov and Don Vandenberg for helpful discussions.
The resources of the Centre de Donn\'ees astronomiques de Strasbourg, Strasbourg, France
(\url{http://cds.u-strasbg.fr}), including the SIMBAD database and the X-Match service, were widely used in this study.
This work has made use of BaSTI, PARSEC, MIST and DSEP web tools.
This work has made use of data from the European Space Agency (ESA) mission {\it Gaia}
(\url{https://www.cosmos.esa.int/gaia}), processed by the {\it Gaia} Data Processing and Analysis Consortium
(DPAC, \url{https://www.cosmos.esa.int/web/gaia/dpac/consortium}).
This study is based on observations made with the NASA/ESA {\it Hubble Space Telescope}.
This publication makes use of data products from the {\it Wide-field Infrared Survey Explorer}, which is a joint project
of the University of California, Los Angeles, and the Jet Propulsion Laboratory/California Institute of Technology.
This work has made use of the Sloan Digital Sky Survey and the United Kingdom Infrared Telescope Infrared Deep Sky Survey.
This publication makes use of data products from the Pan-STARRS1 Surveys (PS1).

\appendix
\section{Some CMDs of NGC\,5904}

\begin{figure}
\includegraphics{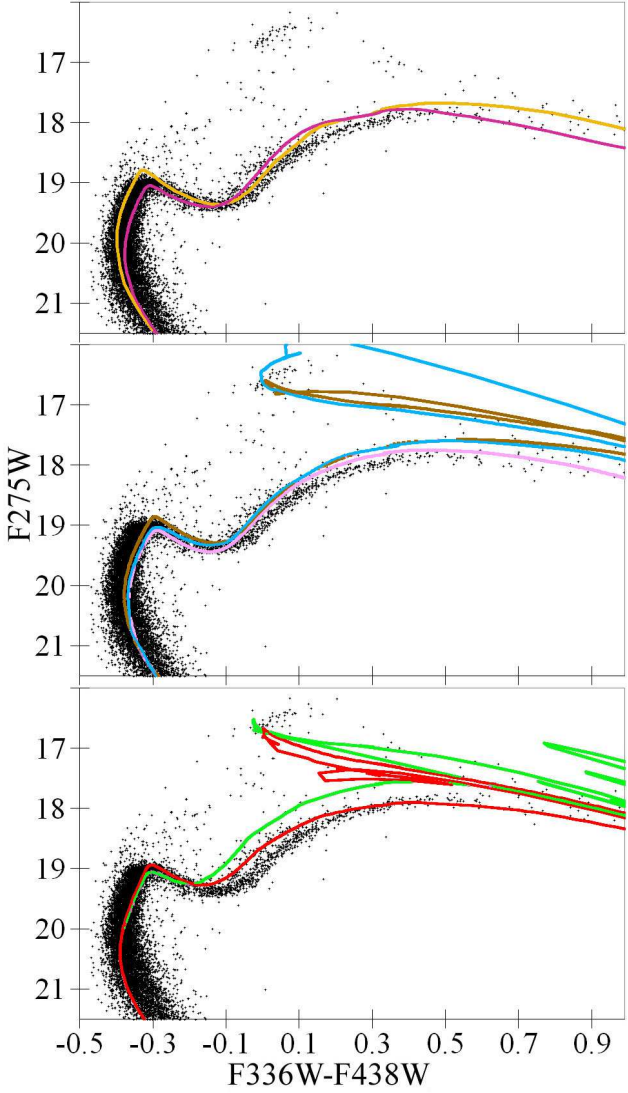}
\caption{$F336W-F438W$ versus $F275W$ CMD of NGC\,5904 with the best-fit parameters from Table~\ref{fit} and
the isochrones from
DSEP solar-scaled (yellow) and DSEP He and $\alpha$ enhanced (magenta) -- upper plot,
old BaSTI solar-scaled (brown), old BaSTI He and $\alpha$ enhanced (light purple), and
new BaSTI solar-scaled (blue) -- middle plot,
PARSEC solar-scaled (green) and MIST solar-scaled (red) -- lower plot.
}
\label{best336_438}
\end{figure}

\begin{figure}
\includegraphics{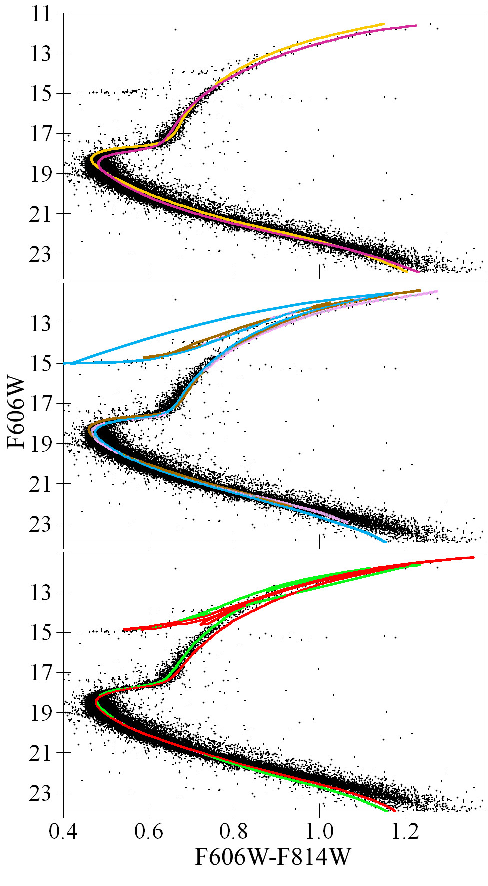}
\caption{The same as Fig.~\ref{best336_438} but for $F606W-F814W$ versus $F606W$.
}
\label{best606_814}
\end{figure}

\begin{figure}
\includegraphics{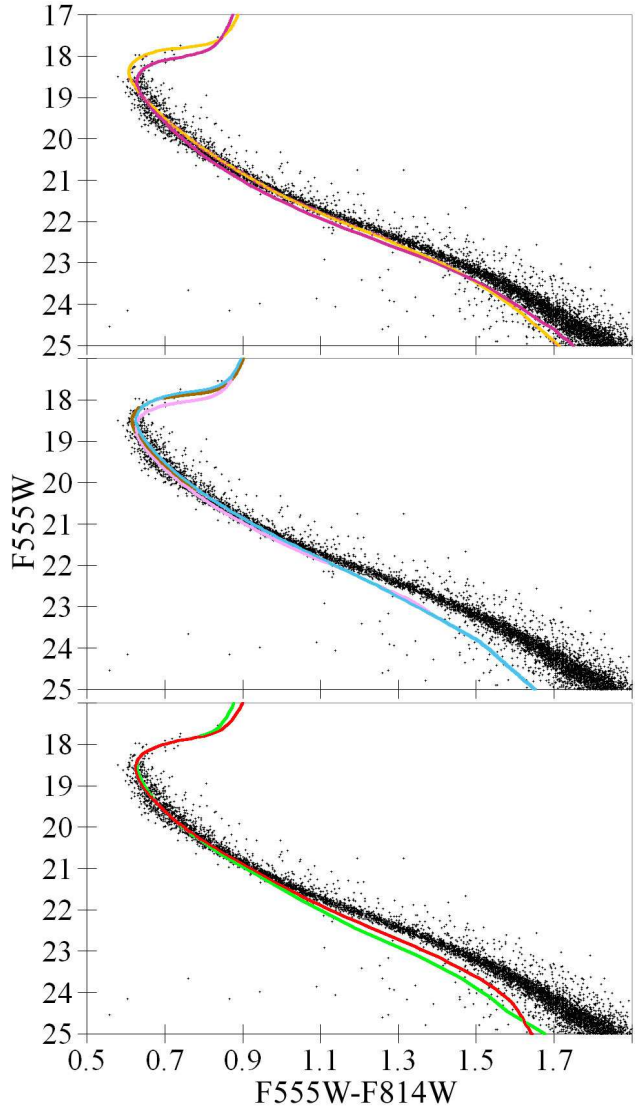}
\caption{The same as Fig.~\ref{best336_438} but for $F555W-F814W$ versus $F555W$.
}
\label{best555_814}
\end{figure}

\begin{figure}
\includegraphics{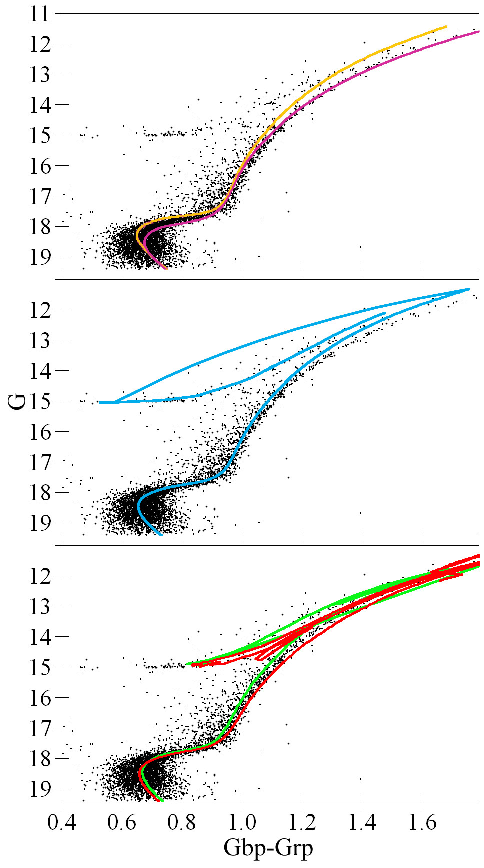}
\caption{The same as Fig.~\ref{best336_438} but for $G_\mathrm{BP}-G_\mathrm{RP}$ versus $G$.
}
\label{bestgbgr}
\end{figure}

\begin{figure}
\includegraphics{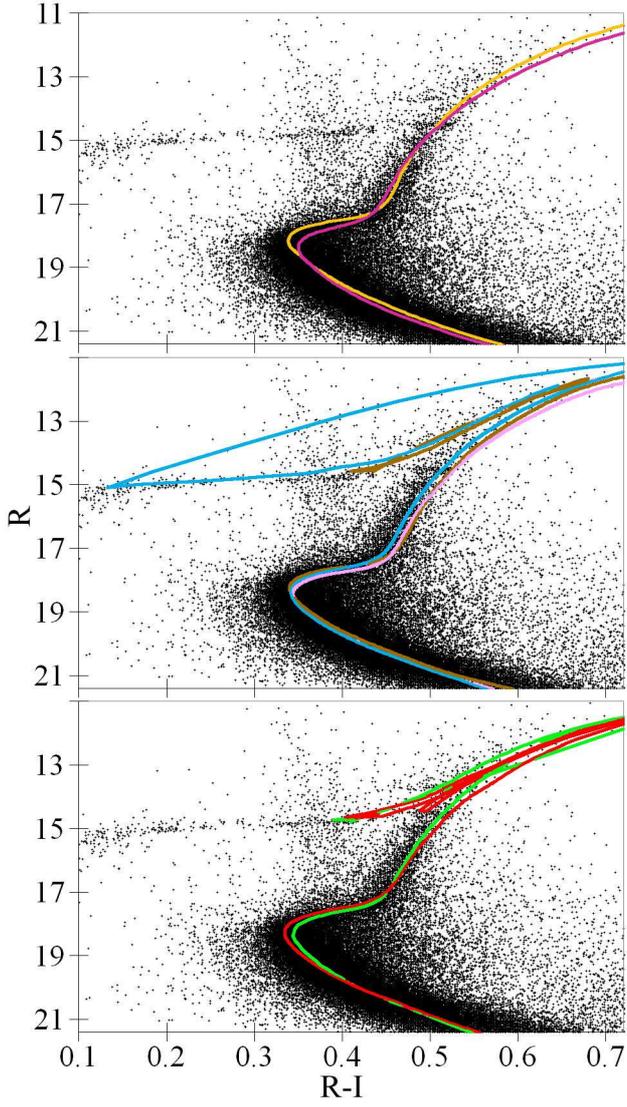}
\caption{The same as Fig.~\ref{best336_438} but for $R-I$ versus $R$.
}
\label{bestri}
\end{figure}

\begin{figure}
\includegraphics{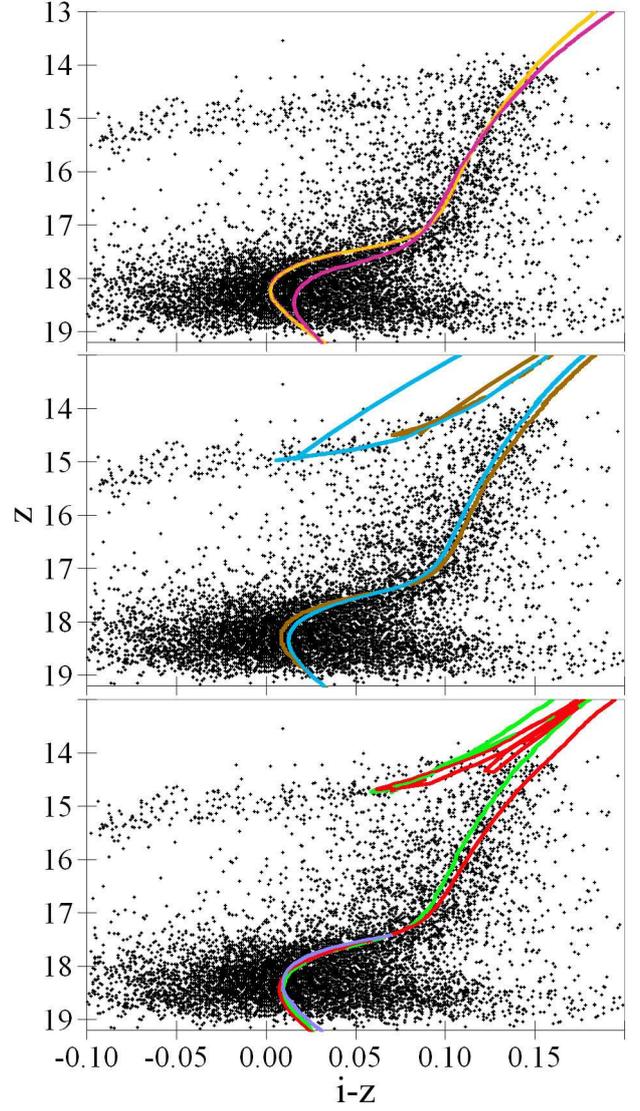}
\caption{The same as Fig.~\ref{best336_438} but for $i-z$ versus $z$.
The additional isochrone from \citetalias{an2009} for the TO and SGB is shown together with the PARSEC and MIST ones 
as the violet curve.
}
\label{bestiz}
\end{figure}

\begin{figure}
\includegraphics{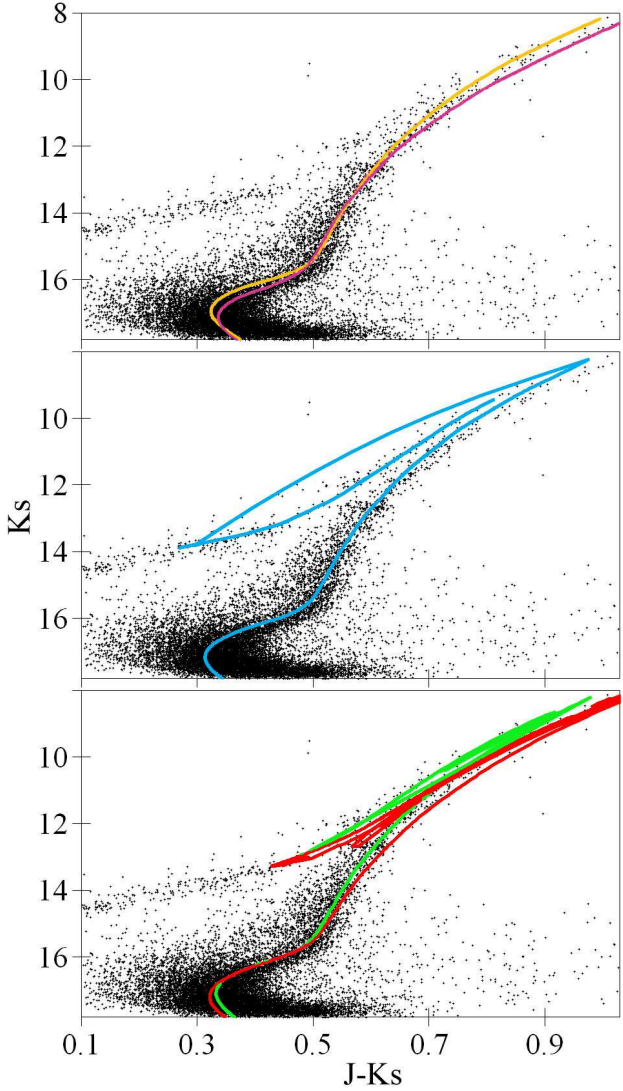}
\caption{The same as Fig.~\ref{best336_438} but for $J-K_s$ versus $K_s$.
}
\label{bestjks}
\end{figure}

\begin{figure}
\includegraphics{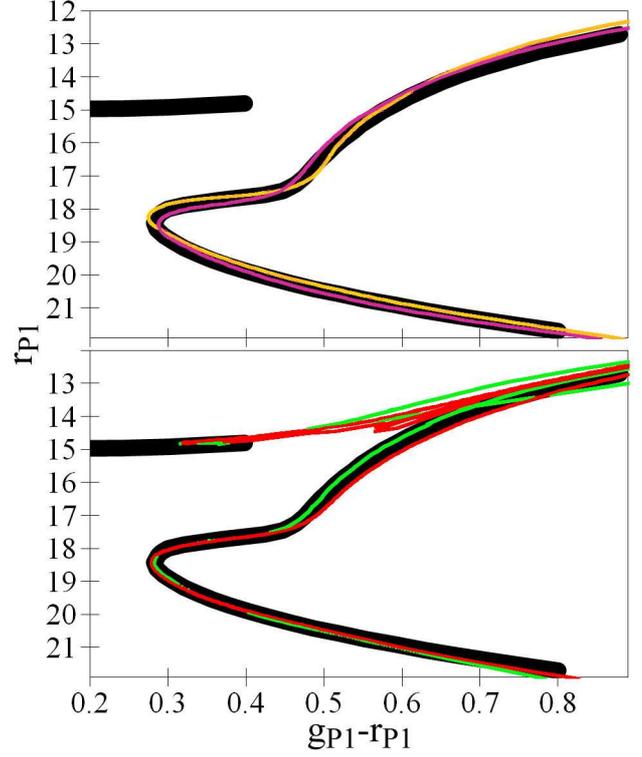}
\caption{Pan-STARRS $g_{P1}-r_{P1}$ versus $r_{P1}$ CMD of NGC\,5904 with the fiducial sequence from
\citet{bernard2014} -- the black thick curve, and the isochrones coloured as in Fig.~\ref{best336_438} with the
best-fit parameters from Table~\ref{fit}.
}
\label{bestpanstarrs}
\end{figure}

\begin{figure}
\includegraphics{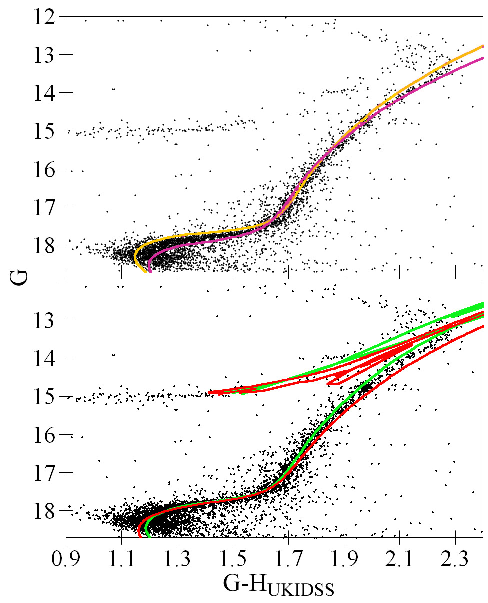}
\caption{The same as Fig.~\ref{best336_438} but for $G-H_\mathrm{UKIDSS}$ versus $G$.
}
\label{bestgh}
\end{figure}

\begin{figure}
\includegraphics{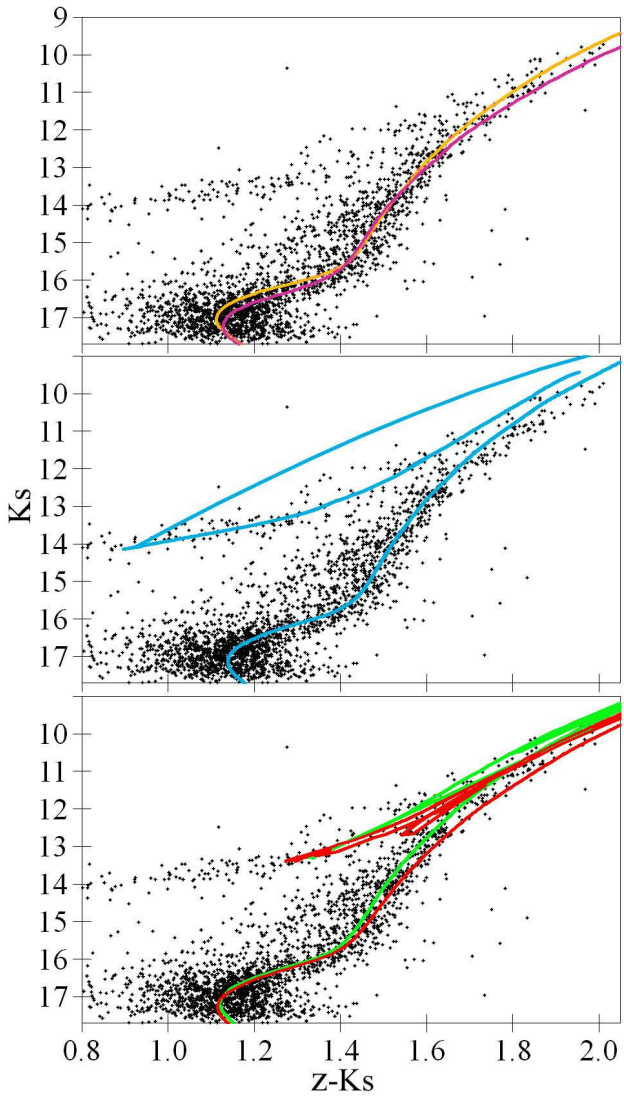}
\caption{The same as Fig.~\ref{best336_438} but for $z-K_s$ versus $K_s$.
}
\label{bestzks}
\end{figure}

\bsp	
\label{lastpage}
\end{document}